\documentclass[]{aa}  
\usepackage[utf8]{inputenx}

\usepackage{graphicx}
\usepackage{fixltx2e}
\usepackage{url}

\usepackage{lscape}

\usepackage{bm}
\usepackage{tikz}
\usetikzlibrary{patterns}
\usepackage{grffile}

\usepackage{color}

\usepackage[varg]{txfonts}
\bibpunct{(}{)}{;}{a}{}{,} % to follow the A&A style

\newcommand{\esothanks}{Based on observations collected at the European Organisation for Astronomical Research in the Southern Hemisphere
under ESO programme 094.A-0205.}

\begin{document} 

\title{The MUSE-Wide Survey: A first catalogue of 831 emission line
  galaxies\thanks{\esothanks}$^{,}$\thanks{At
    \texttt{http://muse-vlt.eu/science/muse-wide-survey} all
    data products described in this paper are available for download.} }

\titlerunning{First catalogue of emission line galaxies in MUSE-Wide}
\authorrunning{E.~C.~Herenz et al.}

\author{
  Edmund~Christian~Herenz\inst{\ref{inst1},}\inst{\ref{inst2}} \and
  Tanya~Urrutia\inst{\ref{inst1}} \and
  Lutz~Wisotzki\inst{\ref{inst1}} \and
  Josephine~Kerutt\inst{\ref{inst1}} \and
  Rikke~Saust\inst{\ref{inst1}} \and
  Maria~Werhahn\inst{\ref{inst1}} \and
  Kasper~Borello~Schmidt\inst{\ref{inst1}} \and
  Joseph~Caruana\inst{\ref{inst1},\ref{inst3},\ref{inst4}} \and
  Catrina~Diener\inst{\ref{inst1},\ref{inst5}} \and
  Roland~Bacon\inst{\ref{inst6}}  \and
  Jarle~Brinchmann\inst{\ref{inst7}} \and
  Joop~Schaye\inst{\ref{inst7}}\and
  Michael~Maseda\inst{\ref{inst7}}\and
  Peter~M.~Weilbacher\inst{\ref{inst1}}
      }  % Michael Maseda

\institute{
  Leibniz-Institut für Astrophysik Potsdam (AIP), An der
  Sternware 16, 14482 Potsdam, Germany \label{inst1}
   \and
 Department of Astronomy, Stockholm University, AlbaNova University Centre, SE-106 91,
 Stockholm, Sweden \label{inst2}
   \and
 Department of Physics, University of Malta, Msida MSD 2080,
 Malta \label{inst3}
   \and
 Institute of Space Sciences \& Astronomy, University of Malta, Msida
 MSD 2080, Malta \label{inst4}
   \and
 Institute of Astronomy, Madingley Road, Cambridge, CB3 0HA,
 UK \label{inst5}
   \and
 CRAL, Observatoire de Lyon, CNRS, Universit\'{e} Lyon 1, 9 avenue
 Charles Andr\'{e}, 69561 Saint Genis-Laval Cedex,
 France \label{inst6}
   \and
 Leiden Observatory, Leiden University, PO Box 9513, 2300 RA Leiden,
 The Netherlands \label{inst7}
}

\abstract{ We present a first instalment of the MUSE-Wide survey,
  covering an area of 22.2 arcmin$^2$ (corresponding to $\sim$20\% of
  the final survey) in the CANDELS/Deep area of the Chandra Deep Field
  South.  We use the MUSE integral field spectrograph at the ESO VLT
  to conduct a full-area spectroscopic mapping at a depth of 1h
  exposure time per 1 arcmin$^2$ pointing. We searched for compact
  emission line objects using our newly developed LSDCat software
  based on a 3-D matched filtering approach, followed by interactive
  classification and redshift measurement of the sources. Our
  catalogue contains 831 distinct emission line galaxies with
  redshifts ranging from 0.04 to 6.  Roughly one third (237) of the
  emission line sources are Lyman $\alpha$ emitting galaxies with
  $3 < z < 6$, only four of which had previously measured
  spectroscopic redshifts.  At lower redshifts 351 galaxies are
  detected primarily by their [\ion{O}{ii}] emission line
  ($0.3 \lesssim z \lesssim 1.5$), 189 by their [\ion{O}{iii}] line
  ($0.21 \lesssim z \lesssim 0.85$), and 46 by their H$\alpha$ line
  ($0.04 \lesssim z \lesssim 0.42$).  Comparing our spectroscopic
  redshifts to photometric redshift estimates from the literature, we
  find excellent agreement for $z<1.5$ with a median $\Delta z$ of
  only $\sim 4 \times 10^{-4}$ and an outlier rate of 6\%, however a
  significant systematic offset of $\Delta z = 0.26$ and an outlier
  rate of 23\% for Ly$\alpha$ emitters at $z>3$. Together with the
  catalogue we also release 1D PSF-weighted extracted spectra and
  small 3D datacubes centred on each of the 831 sources.  }


\maketitle

\section{Introduction}
\label{sec:introduction-musewide-sample}

Most spectroscopic samples of high-redshift galaxies are based on a
photometric pre-selection of targets
\citep[e.g.][]{Noll2004,Vanzella2005,Vanzella2006,Vanzella2008,Popesso2009,Balestra2010,Mallery2012,LeFevre2013,LeFevre2015}.
These surveys have very successfully maximised their spectroscopic
success rates, i.e.\ the fraction of galaxies with scientifically
usable spectra among all targeted objects, by employing photometric
redshift priors.  However, an inevitable conceptual drawback of this
pre-selection approach is that the selection process itself will leave
its imprint on the final sample properties. Moreover, multi-objects
spectrographs have only limited freedom in choosing targets for
simultaneous observation, and even state-of-the-art surveys hardly
ever obtain target sampling rates above 50\%, more often well below
this level. Finally, aperture effects in the preconfigured slit mask
can lead to significant flux losses especially from complex objects
and/or extended emission line regions.

Integral field spectroscopy (IFS) provides an alternative approach
that circumvents many of these problems.  Contiguous areas in the sky
can be mapped instead of targeting individual objects, providing
spectral information of everything within the field of view and within
the sensitivity limits of the observation (see
\citealt{vanBreukelen2005} and \citealt{Adams2011} for pioneering
implementations).  The new panoramic IFS instrument MUSE (Multi Unit
Spectroscopic Explorer) at the ESO Very Large Telescope is a
particularly powerful machine specifically designed to perform blind
surveys for extremely faint high-redshift galaxies
\citep{Bacon2009,Bacon2014,Caillier2014}.  The discovery
potential of MUSE was strikingly demonstrated by a 27 hour integration
in the Hubble Deep Field South \citep{Bacon2015}, where 189
redshifts could be measured inside a single MUSE field of
1~arcmin$^2$. While surveying the sky with MUSE overcomes the above
mentioned limitations, this obviously happens at the expense of
covering only a very small area at a time.

Here we present results from the MUSE-Wide survey, a GTO programme
complementing the ongoing ultra-deep pencil-beam MUSE surveys (Bacon
et al., A\&A submitted). MUSE-Wide trades depth for survey area by
covering many fields with relatively shallow exposures (1h per
field). Nevertheless, the obtained depth already suffices to obtain
source densities (i.e.\ multiplex factors) of several tens of objects
per arcmin$^2$ with useful spectra, at a target sampling rate of
essentially 100\% and with all the benefits of a powerful integral
field unit (IFU). In particular, MUSE features an excellent spatial
sampling at $0\farcs2\times 0\farcs2$ spatial pixels and a spectral
resolution of 2.5\AA\ over one octave in wavelengths from 4750\AA\ to
9350\AA.

A full description of the MUSE-Wide survey strategy will be present in
a forthcoming dedicated publication, accompanying a first general data
release (Urrutia et al., in prep.). Here we only summarise the
currently available material that forms the basis of the present
publication.  The MUSE-Wide survey focuses on areas with extremely deep HST
imaging, with special emphasis on the GOODS-South
\citep{Giavalisco2004} and CANDELS-Deep/CDFS
\citep{Grogin2011,Koekemoer2011} regions, which in addition to the HST
coverage also contains a plethora of multiwavelength data.

This paper describes the outcome of a blind search for emission line
objects in 24 MUSE-Wide fields in the CDFS region, covering a
footprint of 22.2 arcmin$^2$ and yielding a total of 831 galaxies, of
which more than half had no spectroscopic redshift until now.  Besides
the catalogue we also publish object-specific data products suitable
for further investigations.

The structure of the paper is as follows: In
Sect.~\ref{sec:obs-reduc-musewide} we outline the observations and
data reduction. We then describe in
Sect.~\ref{sec:emission-line-source} how we detect, parameterise and
classify emission line sources in our datacubes, including details of
the redshift determination procedure. In
Sect.~\ref{sec:source-catalogue} we present and describe the source
catalogue and data products. Sect.~\ref{sec:char-catal} is dedicated
to a global characterisation of the obtained sample. We present our
conclusions and outlook in Sect.~\ref{sec:musewide-samp-conc}.

\section{Observations and data reduction}
\label{sec:obs-reduc-musewide}

\begin{figure*}
  \centering
  \includegraphics[width=0.8\textwidth]{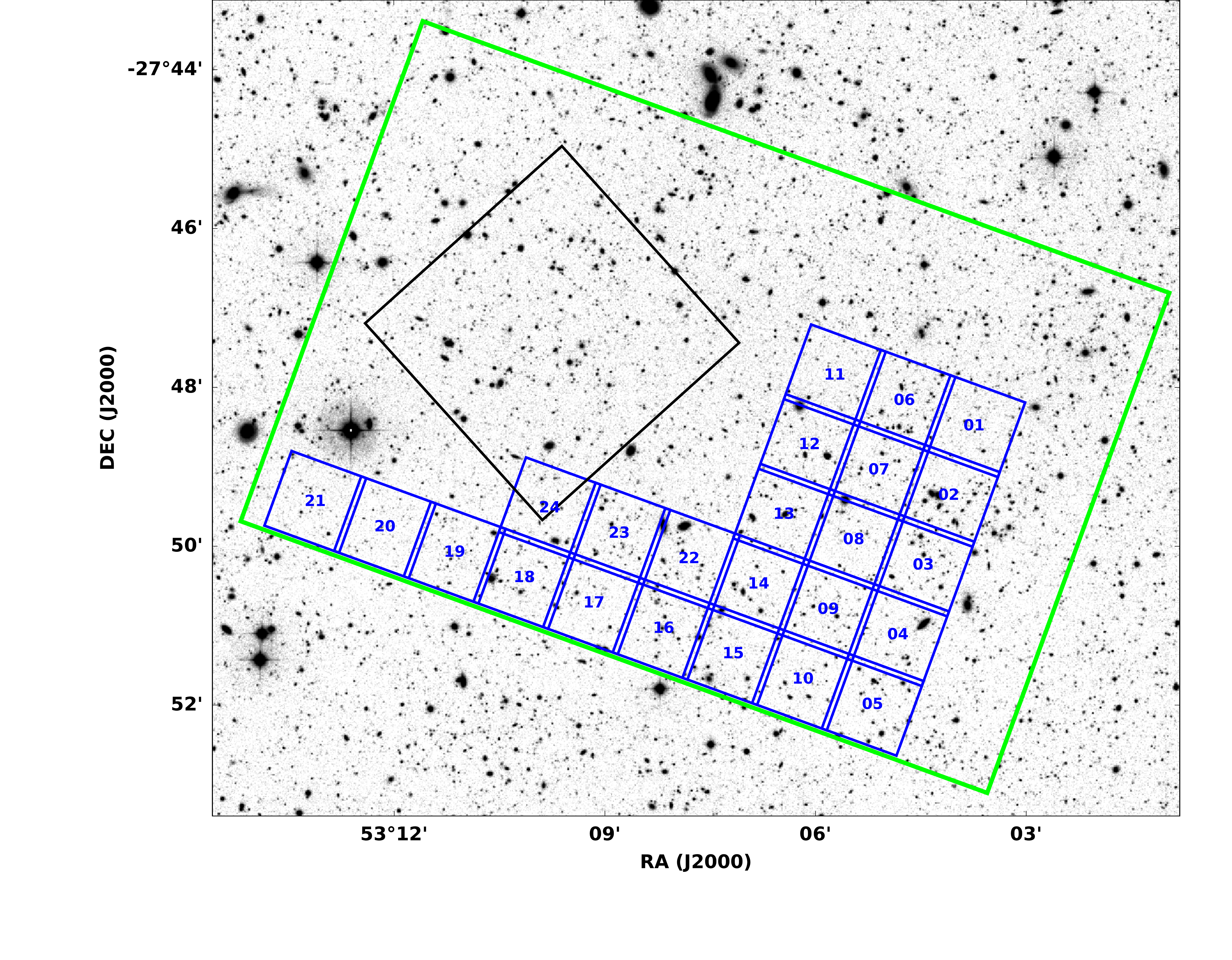} \vspace{-2em}
  \caption{Footprint of the first 24 1\arcmin{}$\times$1\arcmin{}
    pointings of the MUSE-Wide survey in the CANDELS Deep region of
    GOODS-South \citep[overlaid over the R-Band Image from GaBoDS --
    ][]{Erben2005,Hildebrandt2006}. The black square indicates the
    region of the Hubble Ultra Deep Field and the green rectangle
    outlines the CANDELS Deep region.}
  \label{fig:footprint}
\end{figure*}

\begin{table*}
  \caption{Log of Observations}
  \label{tab:musewide_log}
  \centering
  \begin{tabular}{ccclccc} \hline \hline
    Pointing &\multicolumn{2}{c}{Pointing centre}& \multicolumn{1}{c}{Date}        & AG Seeing   & Airmass & Conditions \\
    {}  & $\alpha_\mathrm{J2000}$ & $\delta_\mathrm{J2000}$         &             & [\arcsec{}]   &          &         {} \\ \hline
    MUSE-candels-cdfs-01   & 03$^\mathrm{h}$32$^\mathrm{m}$15.04$^\mathrm{s}$ &-27$^{\circ}$48\arcmin{}29.4\arcsec{}     & 2014-10-20  & 0.86 &  1.087   & photometric / grey\\
    MUSE-candels-cdfs-02   & 03$^\mathrm{h}$32$^\mathrm{m}$16.48$^\mathrm{s}$ &-27$^{\circ}$49\arcmin{}21.5\arcsec{}     & 2014-09-20  & 1.05 &  1.076   & clear / grey\\
    MUSE-candels-cdfs-03   & 03$^\mathrm{h}$32$^\mathrm{m}$17.80$^\mathrm{s}$ &-27$^{\circ}$50\arcmin{}13.6\arcsec{}     & 2014-11-17  & 0.93 &  1.061   & clear / grey\\
    MUSE-candels-cdfs-04   & 03$^\mathrm{h}$32$^\mathrm{m}$19.67$^\mathrm{s}$ &-27$^{\circ}$51\arcmin{}07.1\arcsec{}     & 2014-11-17  & 0.76 &  1.041   & clear / grey \\
    MUSE-candels-cdfs-05   & 03$^\mathrm{h}$32$^\mathrm{m}$20.70$^\mathrm{s}$ &-27$^{\circ}$51\arcmin{}59.8\arcsec{}     & 2014-11-19  & 1.03 &  1.017   & clear / dark\\
    MUSE-candels-cdfs-06   & 03$^\mathrm{h}$32$^\mathrm{m}$18.91$^\mathrm{s}$ &-27$^{\circ}$48\arcmin{}10.7\arcsec{}     & 2014-11-18  & 0.84 &  1.021   & clear / dark \\
    MUSE-candels-cdfs-07   & 03$^\mathrm{h}$32$^\mathrm{m}$20.36$^\mathrm{s}$ &-27$^{\circ}$49\arcmin{}02.6\arcsec{}     & 2014-11-19  & 0.92 &  1.026   & photometric / dark\\ 
    MUSE-candels-cdfs-08   & 03$^\mathrm{h}$32$^\mathrm{m}$21.89$^\mathrm{s}$ &-27$^{\circ}$49\arcmin{}55.3\arcsec{}     & 2014-11-19+20  & 1.00 &  1.034   & photometric / dark \\
    MUSE-candels-cdfs-09   & 03$^\mathrm{h}$32$^\mathrm{m}$23.25$^\mathrm{s}$ &-27$^{\circ}$50\arcmin{}47.9\arcsec{}     & 2014-11-26  & 0.87 &  1.044   & photometric / dark + grey \\
    MUSE-candels-cdfs-10   & 03$^\mathrm{h}$32$^\mathrm{m}$24.68$^\mathrm{s}$ &-27$^{\circ}$51\arcmin{}40.8\arcsec{}     & 2014-11-27  & 0.90 &  1.071   & photometric / grey \\
    MUSE-candels-cdfs-11   & 03$^\mathrm{h}$32$^\mathrm{m}$22.91$^\mathrm{s}$ &-27$^{\circ}$47\arcmin{}50.9\arcsec{}     & 2014-11-28  & 0.95 &  1.104   & photometric / grey\\
    MUSE-candels-cdfs-12   & 03$^\mathrm{h}$32$^\mathrm{m}$24.35$^\mathrm{s}$ &-27$^{\circ}$48\arcmin{}43.2\arcsec{}     & 2014-11-27  & 1.02 &  1.144   & photometric / grey \\
    MUSE-candels-cdfs-13   & 03$^\mathrm{h}$32$^\mathrm{m}$25.88$^\mathrm{s}$ &-27$^{\circ}$49\arcmin{}36.2\arcsec{}     & 2014-11-27  & 1.07 &  1.193   & photometric / dark \\
    MUSE-candels-cdfs-14   & 03$^\mathrm{h}$32$^\mathrm{m}$27.21$^\mathrm{s}$ &-27$^{\circ}$50\arcmin{}28.9\arcsec{}     & 2014-11-28  & 0.88 &  1.229   & photometric / grey \\
    MUSE-candels-cdfs-15   & 03$^\mathrm{h}$32$^\mathrm{m}$28.66$^\mathrm{s}$ &-27$^{\circ}$51\arcmin{}21.5\arcsec{}     & 2014-12-25  & 0.83 &  1.241   & photometric / grey\\
    MUSE-candels-cdfs-16   & 03$^\mathrm{h}$32$^\mathrm{m}$32.62$^\mathrm{s}$ &-27$^{\circ}$51\arcmin{}02.1\arcsec{}     & 2014-11-28  & 0.83 &  1.228   & photometric / grey \\
    MUSE-candels-cdfs-17   & 03$^\mathrm{h}$32$^\mathrm{m}$36.60$^\mathrm{s}$ &-27$^{\circ}$50\arcmin{}43.8\arcsec{}     & 2014-12-23  & 0.80 &  1.191   & clear / dark \\
    MUSE-candels-cdfs-18   & 03$^\mathrm{h}$32$^\mathrm{m}$40.58$^\mathrm{s}$ &-27$^{\circ}$50\arcmin{}24.3\arcsec{}     & 2014-12-21  & 0.89 &  1.191   & photometric / dark \\
    MUSE-candels-cdfs-19   & 03$^\mathrm{h}$32$^\mathrm{m}$44.60$^\mathrm{s}$ &-27$^{\circ}$50\arcmin{}04.6\arcsec{}     & 2014-12-21  & 0.82 &  1.223   & photometric / dark \\
    MUSE-candels-cdfs-20   & 03$^\mathrm{h}$32$^\mathrm{m}$48.61$^\mathrm{s}$ &-27$^{\circ}$49\arcmin{}46.0\arcsec{}     & 2014-12-23  & 0.82 &  1.288   & clear / dark \\
    MUSE-candels-cdfs-21   & 03$^\mathrm{h}$32$^\mathrm{m}$52.54$^\mathrm{s}$ &-27$^{\circ}$49\arcmin{}26.1\arcsec{}     & 2014-12-23  & 0.72 &  1.388   & clear / dark \\
    MUSE-candels-cdfs-22   & 03$^\mathrm{h}$32$^\mathrm{m}$31.19$^\mathrm{s}$ &-27$^{\circ}$50\arcmin{}09.8\arcsec{}     & 2014-12-22  & 0.79 &  1.338   & clear / dark\\
    MUSE-candels-cdfs-23   & 03$^\mathrm{h}$32$^\mathrm{m}$35.23$^\mathrm{s}$ &-27$^{\circ}$49\arcmin{}50.3\arcsec{}     & 2014-12-24  & 0.86 &  1.266   & photometric / dark\\
    MUSE-candels-cdfs-24   & 03$^\mathrm{h}$32$^\mathrm{m}$39.14$^\mathrm{s}$ &-27$^{\circ}$49\arcmin{}31.5\arcsec{}     & 2014-12-26  & 0.81 &  1.168   & photometric / grey\\ \hline \hline
  \end{tabular}
  \tablefoot{The integration time for each exposure was 3600 seconds. Autoguider seeing (AG seeing) and airmass values refer to the average over all
    four individual exposures per pointing.}
\end{table*}

Our current dataset is based on the analysis of 24 adjacent
1\arcmin{}$\times$1\arcmin{} MUSE pointings in the CANDELS Deep region
of the GOODS-South field obtained during the first semester of MUSE
guaranteed time observations.  Observations were carried out in grey
and dark time under photometric and clear conditions from September to
December 2014 (ESO programme 094.A-0205, PI: Lutz Wisotzki). In
Figure~\ref{fig:footprint} we show the footprint of the survey area.  We
matched the position angle of our pointings to the 70$^\circ$ position
angle (east of north) of the CANDELS Deep region (indicated by the
green box in Figure~\ref{fig:footprint}). In
Table~\ref{tab:musewide_log} we provide a log of our observations.
Standard star exposures were taken at the beginning and at the end of
each night.

We integrated 1 hour on each pointing.  Each integration was split
into 4 exposures of 15 minutes.  In between exposures small dither
offsets (typically smaller than 1\arcsec{}) were applied and the
spectrograph was rotated by 90 degrees.  This procedure, which is
recommended in the MUSE User's
manual\footnote{\url{http://www.eso.org/sci/facilities/paranal/instruments/muse/doc.html}},
ensures that patterns of the 24 individual spectrographs and their
image slicers are averaged out.  With the exception of pointings 08,
09, 12, and 16 all exposures for a pointing were obtained in immediate
succession.  Pointing 08 was split into two exposure sequences in two
subsequent nights, while pointings 09, 12, and 16 where taken at
different times during a night.  Adjacent pointings have an overlap of
4\arcsec{}.  Taking this overlap and the exact geometry of the MUSE
field of view into account, the total area exposed with MUSE in the 24
pointings taken in the first cycle of the MUSE-Wide survey is
22.2\,arcmin$^2$.

For the reduction of the individual pointings we used version 1.0 of
the MUSE data reduction system\footnote{Available from ESO via
  \url{http://www.eso.org/sci/software/pipelines/muse/muse-pipe-recipes.html}.}
\citep[][and in prep.]{Weilbacher2014}, in combination with custom
developed \texttt{python}\footnote{\url{http://www.python.org}}
routines and the ZAP tool presented in \cite{Soto2016}.  An in-depth description
and validation of the data reduction procedure will be given in the
publication complementing the full data release (Urrutia et al., in
prep.), in the following we only provide a brief overview.

We used the set of calibration exposures taken closest in time to the
actual observations to create master biases, master flats, dispersion
solutions, and trace tables.  For the illumination correction, we
always chose the illumination frames that were taken before the
science observation.  Using the standard-star exposures we constructed
response curves for flux-calibration.  We applied these calibration
products with the pipeline routine \texttt{muse\_scibasic} to all 24
spectrographs CCD images belonging to one science exposure.  The
result of this process are so-called pixel tables for each exposure
containing calibrated flux values, errors, wavelengths, and
information on their location on the sky.

It is known that the current version of the MUSE pipeline
sky-subtraction routine \citep{Streicher2011} leaves significant
systematic residuals that hamper the detection of faint object signals
\citep{Soto2016}.  For this reason we developed and applied our own
sky-subtraction routine that works on the pixel tables.  We found,
that this procedure, in combination with the ZAP tool by
\cite{Soto2016}, provided a more optimal result compared to using only
ZAP.  Our procedure will be detailed in Urrutia et al., (in prep.).
Finally, we applied the self-calibration method described in Sect.~3.1
of \cite{Bacon2015} to remove systematic mean zero-flux level offsets
between slices\footnote{The self-calibration procedure is part of the
  \emph{Muse Python Data Analysis Framework} (MPDAF,
  \citealt{Bacon2016}, see also \citealt{Conseil2016}).  MPDAF is
  available at \url{http://mpdaf.readthedocs.io/en/latest}.}.

For each exposure we then used the pipeline routine
\texttt{muse\_scipost} that resamples the pixtables into datacubes and
propagates errors into a corresponding variance datacube. The
datacubes were corrected for differential atmospheric refraction using
the formula by \cite{Filippenko1982}.  Remaining sky-subtraction
residuals after the application of our own routine were then purged
from the cubes using the ZAP-software of \cite{Soto2016}.  We then
created white-light images from these datacubes by summation over the
spectral axis.  By performing 2D Gaussian fits to compact objects
within those images (compact galaxies, or, when available, stars) we
determined a reference registration for each exposure datacube.  Using
these determined offsets for the individual cubes, we ran
\texttt{muse\_scipost} again, to resample all exposures onto a common
world-coordinate system grid. We created a combined datacube for each
pointing by averaging the corresponding four exposures with rejection
of 3$\sigma$ outliers.

In an additional postprocessing step we adjusted the zero level for
all layers. This was needed since none of the preceding reduction
steps (including the sky subtraction) actually enforces a spatially
coherent zero background level. We obtained this zero level correction
by masking out all continuum sources in a wavelength-collapsed
white-light image of the field and then computing the average of the
remaining unmasked spaxels in each layer. While these corrections were
found to be small except very close to strong night sky lines, with
median values around $2\times 10^{-22}$ erg s$^{-1}$ cm$^{-2}$
\AA$^{-1}$, as systematic offsets they are nevertheless potentially
significant for large aperture integrations. We subtracted the
corrections layer by layer from the combined cubes and thus obtained
the final flux datacube.

After the procedure described above we have 24 datacubes that contain
$\approx$3.5$\times10^8$ exposed volume-pixels (so called voxels).
The spatial sampling is 0.2\arcsec{}$\times$0.2\arcsec{}, the spectral
sampling is 1.25\AA{}, and the wavelength range extends from 4750\AA{}
to 9350\AA{}.  The wavelength axis is given in air wavelengths in the
barycentric reference frame.  For each cube the astrometry is such
that the east-west and south-north axes are parallel to the spatial
coordinate axes.  Complementary to the flux datacubes we produced a
final variance datacube by propagating the individual variance values.
We also created exposure-map cubes, where we track the number of
individual single exposures that went into a voxel of the final
datacube.  The datacube, variance cube and exposure-map cubes are
stored in separate header and data units (HDUs) of a single FITS file
\citep{Pence2010} taking 5 GB of disk space for each pointing.

\section{Emission line source detection and classification}
\label{sec:emission-line-source}

\subsection{Detection and parameterisation of emission line source
  candidates}
\label{sec:detection}

On each of the 24 datacubes we performed the following tasks to build a
catalogue of emission line source candidates:
\begin{enumerate}
\item \label{item:mw1} Empirical estimation and correction of the MUSE
  pipeline propagated variance cubes.
\item \label{item:mw2} Removal of source continua from the
  datacube.
\item \label{item:mw3} Cross-correlation of the datacube with a 3D
  matched filter for compact emission line sources.
\item \label{item:mw4} Thresholding and cataloguing of emission line
  source candidates.
\item \label{item:mw5} Position, size, and flux measurements of the
  emission line source candidates.
\end{enumerate}
For tasks \ref{item:mw3} to \ref{item:mw5} we have developed the
emission \emph{Line Source Detection and Cataloguing Tool}
\texttt{LSDCat}\footnote{\texttt{LSDCat} is available via the
  Astrophysics Source Code Library: \url{http://www.ascl.net/1612.002}
  \citep{Herenz2016a}.}.  In the following subsections we briefly
describe the above steps.  For an in-depth description of
\texttt{LSDCat} we refer to \cite{Herenz2017} (in the following
HW17).

\subsubsection{Empirical correction of the MUSE pipeline propagated noise}
\label{sec:effective-noise}

\begin{figure*}
  \centering
  \includegraphics[width=\textwidth]{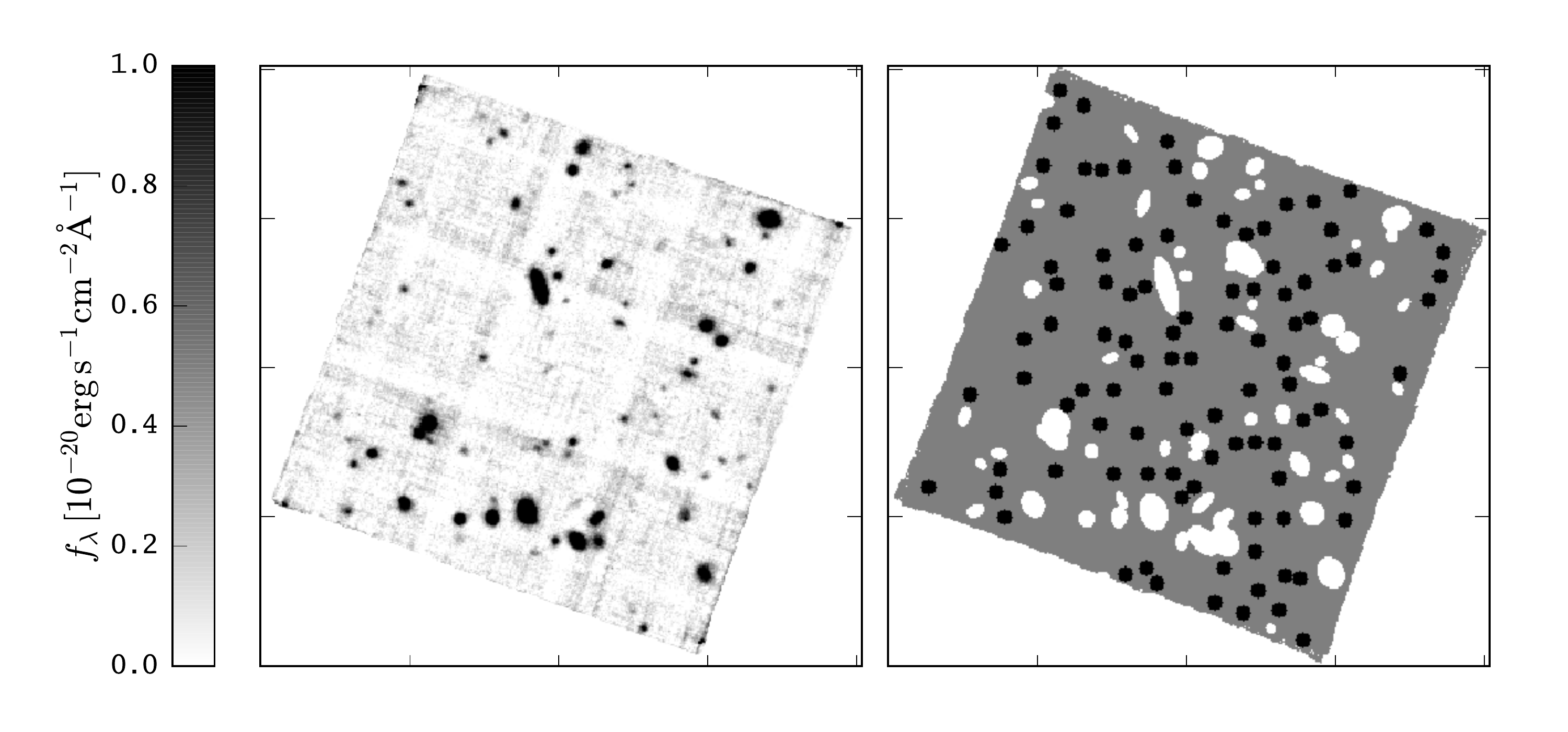}\vspace{-1.5em}
  \includegraphics[width=\textwidth]{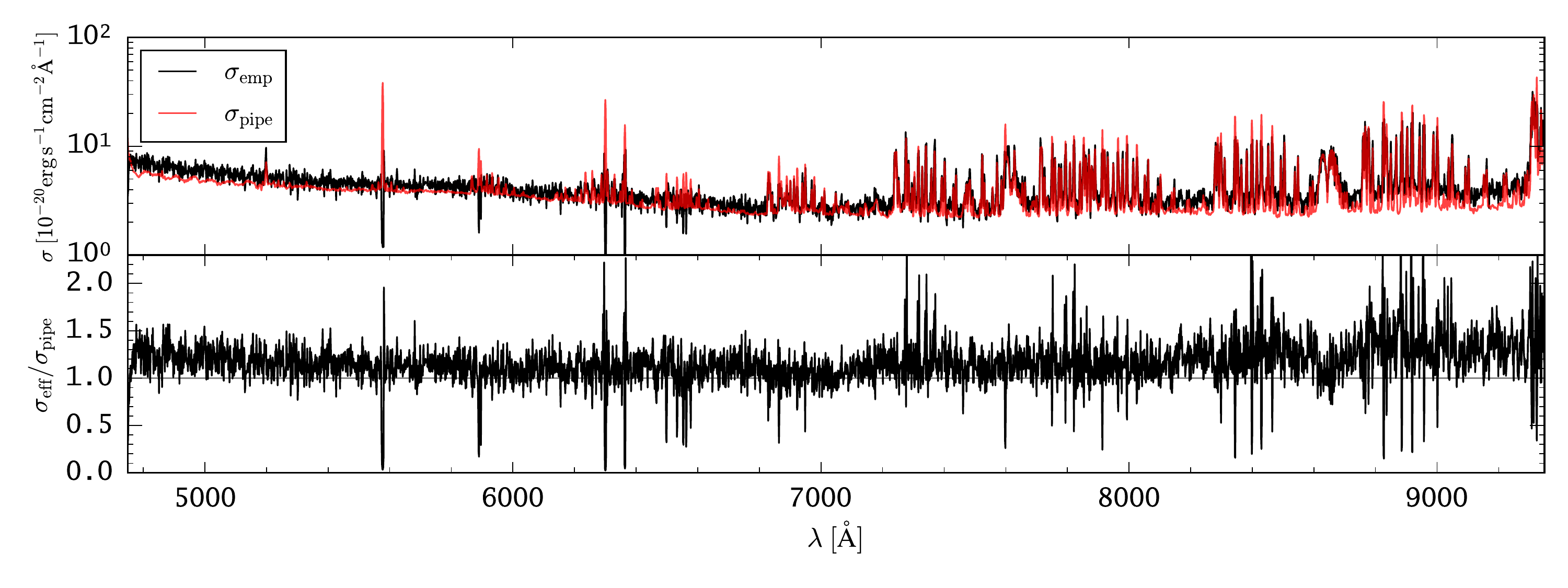}\vspace{-1em}
  \caption{Illustration of the empirical noise calculation procedure.
    Shown is the case in the MUSE-Wide pointing
    MUSE-candels-cdfs-06. The \emph{top left panel} shows a
    white-light image created by summing over all spectral layers and
    subsequent division by the spectral range.  In the \emph{top right
      panel} we show the 100 random 2\arcsec{} diameter apertures
    (black) and the avoided regions (white) because of the presence of
    continuum bright objects \citep[sources with
    $\mathrm{m}_\mathrm{F814} < 25$\,mag in ][]{Guo2013}.  In the
    \emph{bottom panels} we compare the width of the distribution of
    the flux values extracted in the 100 apertures for each spectral
    layer $\sigma_\mathrm{emp}$ (black curve) normalised to one spectral pixel to
    the corresponding aperture average from the pipeline produced
    variance cube $\sigma_\mathrm{pipe}$ (red curve).}
  \label{fig:mweffnoise}
\end{figure*}

Both for emission line source detection and flux measurements we
require an accurate characterisation of the noise in our datacubes.
However, we expect that the variance cubes provided by the pipeline
underestimate the true variance.  First, because the pipeline propagates
variances from the CCD level through the various reduction steps, thus
these formal variance values do not account for hidden (non-Gaussian)
systematics such as imperfect flat-fielding or imperfect
sky-subtraction.  Second, in the resampling process carried out by
the MUSE pipeline co-variance terms are neglected. 

In order to quantify and correct the formal variances produced by the
pipeline, we determined an empirical variance estimate from the flux
datacubes by analysing the pixel statistics for each datacube layer in
regions of blank sky.  However, direct pixel-to-pixel noise statistics
as such would also be biased towards lower values since also here
covariance terms introduced in the resampling process would be
neglected.  To overcome this issue we placed circular apertures with a
2\arcsec{} diameter (10 pixels) at random positions in each pointing.
By prohibiting these apertures to overlap with
$\mathrm{m}_\mathrm{F814} < 25$\,mag sources from the \cite{Guo2013}
CANDELS GOODS-S photometric catalogue we ensured that blank sky was
sampled.  The width of the distribution of average pixel values in
those apertures, characterised by its standard deviation, was then
used as an estimate of the noise for each spectral layer.  Due to the
small FoV of MUSE we were limited to use only 100 apertures.  Hence,
we estimated the width of the distribution for each layer via
$\sigma_\mathrm{emp} = (q_{75} - q_{25}) / (2 \sqrt{2}
\mathrm{erf}^{-1}(1/2)) \approx 0.7413 \, (q_{75} - q_{25})$, where
$\mathrm{erf}^{-1}$ is the inverse error function and
$q_{75} - q_{25}$ is the interquartile range of the distribution
\citep[e.g., Sect. 3.2.2 in ][]{Ivezic2014}.  We then compared the so
obtained empirical noise values $\sigma_\mathrm{emp}^2$ to an average
value $\sigma_\mathrm{pipe}^2$ of the pipeline propagated variance in
each layer -- this process is illustrated in
Figure~\ref{fig:mweffnoise}.  As can be seen the ratio
$\sigma_\mathrm{emp}/\sigma_{\mathrm{pipe}}$ is greater than one in
almost all datacube layers.  For all cubes we found typically
$\sigma_\mathrm{emp} = 1.15 - 1.20 \,\sigma_\mathrm{pipe}$.  This is
consistent with an empirical noise estimate made for MUSE datacubes
obtained with a similar observing strategy \citep[][their
Sect. 3.3]{Borisova2016}.  However, as can be also seen in
Figure~\ref{fig:mweffnoise} there is a small number of layers where
the ratio $\sigma_\mathrm{emp}/\sigma_{\mathrm{pipe}}$ is less than
one.  The affected layers are in the cores of sky emission lines where
they suffer from over-subtraction of the high-frequency noise by the
ZAP routine.

In order to correct the pipeline propagated variance estimate, we
replace the values in the variance cubes with the empirical noise
estimate $\sigma_\mathrm{emp}^2$.  In layers where
$\sigma_\mathrm{emp}/\sigma_\mathrm{pipe} < 1$ we use the average
value from the pipeline.  Due to the dither- and rotation pattern
typically $\lesssim 10$\% of voxels of the datacube have not
contributions from all four exposure.  These voxels are mostly on the
border of the FoV.  To adjust the noise estimates for those voxels we
rescale their $\sigma_\mathrm{emp}^2$ with $4/N_\mathrm{exp}$, where
$N_\mathrm{exp}$ is the number of exposures contributing to that
voxel.

There are two caveats with this empirical noise estimate.  First, due
to the small number of apertures our noise estimate is itself noisy.
This can be seen when comparing the smooth
$\sigma_\mathrm{pipe}$-curve to the noisier
$\sigma_\mathrm{emp}$-curve in Figure~\ref{fig:mweffnoise}.  Second,
our estimate does not take correlated noise in the spectral direction
into account.  We will address these shortcomings of the described
empirical noise estimate in future releases of the MUSE-Wide data
(Urrutia et al., in prep.).  Nevertheless, compared to the MUSE
pipeline propagated variances that underestimate the true variance our
empirical estimation approach is an improvement.

\subsubsection{Removal of source continua}
\label{sec:remov-cont-sign}

\begin{figure*}
  \centering
  \includegraphics[width=\textwidth]{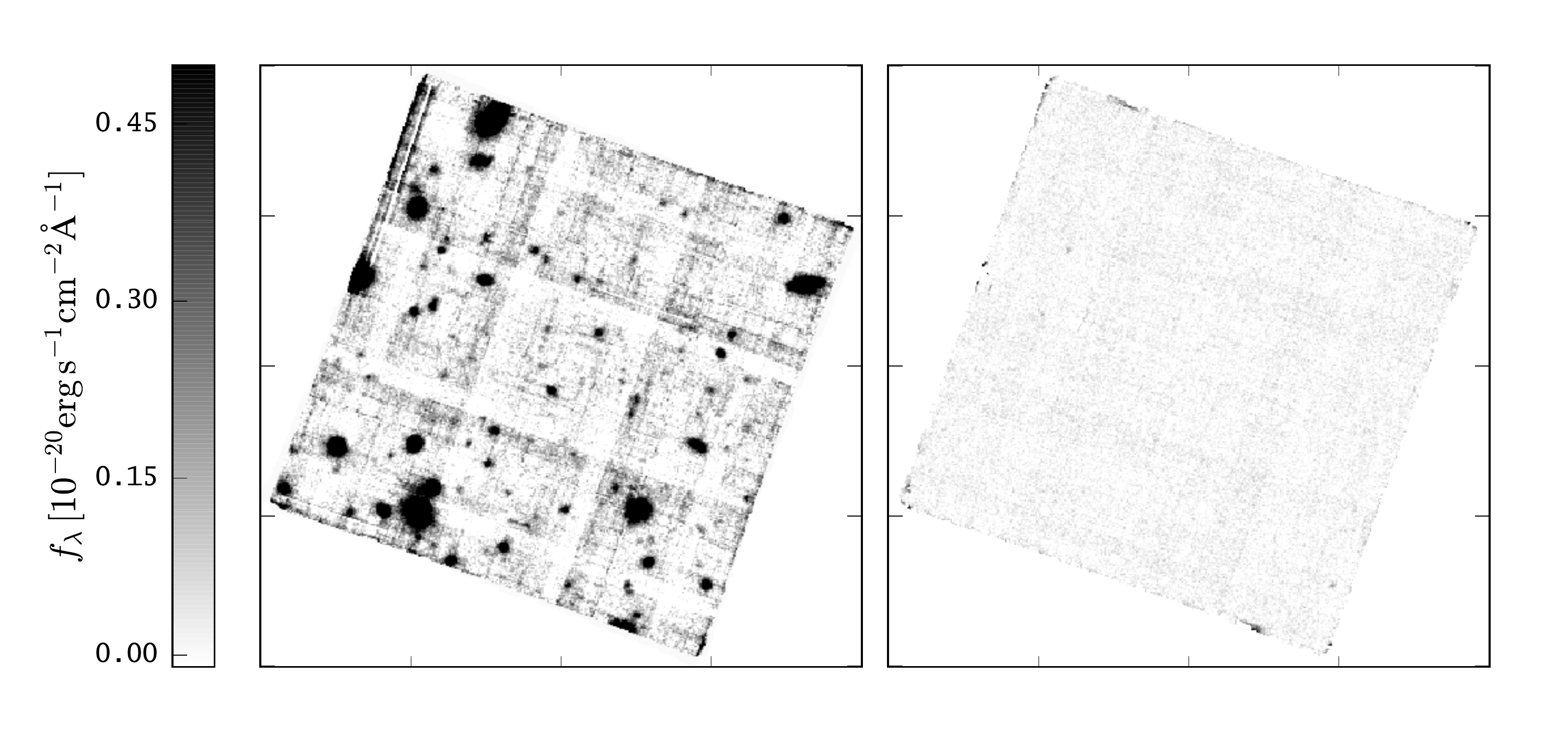}
  \vspace{-1.2em}
  \caption{Example illustrating of the effectiveness of subtracting a
    running median in the spectral direction from the datacube to
    remove signal from bright continuum sources.  Shown is the
    MUSE-Wide pointing MUSE-candels-cdfs-15.  \emph{Left}: White-light
    image created by summing all spectral layers and subsequent
    division by the spectral range. \emph{Right}: White-light image of
    the median filter subtracted version of the same cube illustrating
    the effectiveness of this continuum subtraction method.  }
  \label{fig:medfilsubex}
\end{figure*}

The source detection algorithm in \texttt{LSDCat} searches for
emission line signals and implicitly assumes that significant source
continua are subtracted from the datacube.  We achieved this by
subtracting a 151 pixel wide running median in the spectral direction
from the datacube.  The full width of the median filter was chosen to
be 151 spectral layers (188.75\AA{}), which is much broader than the
width of the targeted emission lines and narrow enough to robustly
subtract slowly varying continua.  The remaining residuals in the
datacube are either real emission lines or residual features from
continua varying at high frequencies (e.g. cold stars).  We
demonstrate the effectiveness of our median-filter subtraction method
in Fig.~\ref{fig:medfilsubex}.

\subsubsection{Cross-correlation with matched filter}
\label{sec:cross-corr-with}

\begin{figure}
  \centering
  \includegraphics[width=0.5\textwidth,trim=0 0 0 0.3cm,clip=true]{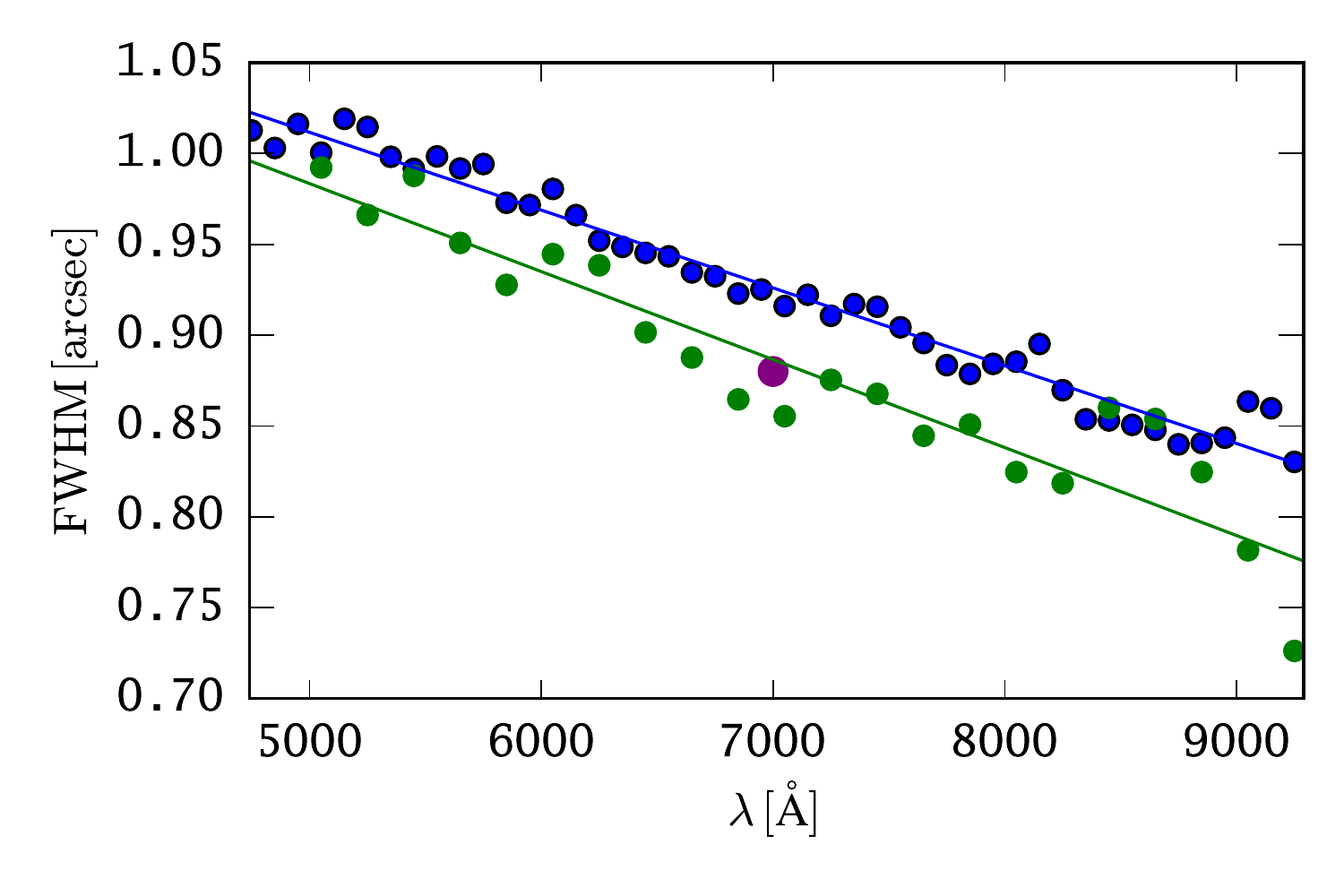}\vspace{-1em}
  \caption{Illustration of the determination of the wavelength
    dependence of the FWHM of the seeing PSF. As an example we show the
    results for the MUSE-Wide pointing
    MUSE-candels-cdfs-14. \emph{Blue points} show the FWHM values
    obtained from fitting a 2D Gaussian to images of a star within the
    datacube.  The used images were created by summing over 100\,\AA{}
    along the spectral axis. \emph{Green points} show the results from
    minimising the $\chi^2$ difference between MUSE images of several
    compact galaxies within a pointing to 2D Gaussian-convolved and to
    MUSE resolution resampled HST images of those galaxies.  For each
    image the FWHM of the 2D Gaussian kernel minimising $\chi^2$ is
    displayed.  Here the used images are created by summing over
    200\,\AA{} along the spectral axis.  The green and the blue lines
    are the linear fits
    $\mathrm{FWHM}(\lambda) = p_0 + p_1 (\lambda -
    7000\,\mathrm{\AA{}})$ to the individual data points of the FWHM
    determination using a star or several compact galaxies,
    respectively.  The purple point at $7000\,\mathrm{\AA{}}$ is the
    mean value inferred from the VLT auto guider probe (AG seeing)
    averaged over all four exposures.  As the green points and line
    are closer to the AG seeing value, we chose this fit to describe
    the $\mathrm{FWHM}(\lambda)$ dependence.}
  \label{fig:psffitsjosie}
\end{figure}
\begin{table}
\centering
    \centering
    \caption{
      Coefficients $p_0$ and $p_1$ of
      $\mathrm{FWHM}(\lambda) = p_0 + p_1 (\lambda -
      7000\,\mathrm{\AA{}})$ that describe the wavelength dependence of
      the PSF FWHM in each of the datacubes. These coefficients are
      used as  input parameters in \texttt{LSDCat} for the spatial filtering.}
    \label{tab:psffwhmmusewide}
    \begin{tabular}{cccl} \hline \hline
      Pointing & $p_0$ & $p_1$ & Method \\
      No.  & [\arcsec{}] & [$10^{-5}$\,\arcsec{}/\AA{}] & {} \\ \hline
      01    & 0.836       & -4.429 & fit to ID 10548, $m_\mathrm{F814W}=22.18$ \\
      02    & 0.940       & -3.182 & compact galaxies, no star \\
      03    & 0.944       & -4.460 & fit to ID  8374, $m_\mathrm{F814W}=19.39$ \\
      04    & 0.747       & -4.219 & compact galaxies, no star \\
      05    & 1.026       & -3.003 & compact galaxies, no star \\
      06    & 0.835       & -4.332 & compact galaxies, no star \\
      07    & 0.935       & -3.966 & compact galaxies, no star \\
      08    & 0.991       & -5.007 & compact galaxies \\
      09    & 0.833       & -8.069 & fit to ID 68879, $m_\mathrm{F814W}=19.73$ \\
      10    & 0.890       & -3.051 & compact galaxies \\
      11    & 0.989       & -3.771 & compact galaxies \\
      12    & 1.020       & -4.123 & compact galaxies, no star \\
      13    & 1.063       & -5.285 & compact galaxies \\
      14    & 0.884       & -4.844 & compact galaxies \\
      15    & 0.702       & -4.441 & fit to ID 5744, $m_\mathrm{F814W}=20.3$ \\
      16    & 0.859       & -3.784 & fit to ID 6475, $m_\mathrm{F814W}=18.86$ \\
      17    & 0.780       & -3.534 & compact galaxies \\
      18    & 0.929       & -3.479 & compact galaxies, no star \\
      19    & 0.814       & -3.524 & compact galaxies, no star \\
      20    & 0.713       & -5.196 & compact galaxies, no star \\
      21    & 0.836       & -4.255 & fit to ID 9801, $m_\mathrm{F814W}=21.92$ \\
      22    & 0.788       & -3.253 & fit to ID 7813, $m_\mathrm{F814W}=20.23$ \\
      23    & 0.777       & -3.019 & compact galaxies, no star \\
      24    & 0.728       & -4.232 & compact galaxies, no star \\
      \hline
    \end{tabular}
    \tablefoot{In the method column we give the ID and F814 magnitude
      from the \cite{Guo2013} catalogue when the polynomial
      coefficients were derived using this star.  Otherwise we
      indicate with ``compact galaxies'', that the minimisation
      utilising compact galaxies provided a $p_0$ value closer to the
      AG Seeing and we therefore used $p_0$ and $p_1$ from this
      method.  With ``compact galaxies, no star'' we indicate that no
      sufficiently bright star was present within the datacube, and we
      thus had to rely on the minimisation scheme utilising compact
      galaxies.}
\end{table}

The detection algorithm of \texttt{LSDCat} is based on
matched filtering, an operation that maximises the S/N ratio of emission
lines within the datacube
\citep[e.g.][]{Schwartz1975,Das1991,Bertin2001,Zackay2015,Vio2016}.
To this aim the algorithm transforms the input datacube  by convolving it
with a three-dimensional template that matches the expected signal of
an emission line in the datacube. The 3D convolution is performed as
two successive convolutions, a 2D convolution in each spectral layer
and a 1D convolution in the spectral direction for each spaxel.

As the template for the 2D convolution in each spectral layer we use
the circular Gaussian profile option of \texttt{LSDCat}.  For the
matched filtering this template provides a reasonably good
approximation of the seeing induced point spread function (PSF).  The
width of the PSF, typically given as the full width at half maximum
(FWHM), depends on wavelength \citep[e.g.,][]{Hickson2014}.  The
input-parameters for \texttt{LSDCat} describing this width and its
wavelength dependence need to be supplied as the coefficients $p_0$
and $p_1$ of the linear function
$\mathrm{FWHM}(\lambda)[\arcsec] = p_0 + p_1(\lambda -
7000\mathrm{\AA{}})$ that provides an acceptable approximation over
the wavelength range under consideration here.  In principle, the
determination of the $\mathrm{FWHM}(\lambda)$ dependence, and thus the
optimal coefficients in $p_0$ and $p_1$, can be achieved by fitting a
2D Gaussian function to a reasonably bright star in each spectral
layer of the datacube.  However, by choice the CANDELS fields are
devoid of bright stars.  Hence, for numerous of our pointings
selecting objects from the \cite{Guo2013} CANDELS GOODS-S photometric
catalogue with $\mathtt{CLASS\_STAR} > 0.95$, i.e. objects that are
likely stars, results only in objects with
$\mathrm{m}_\mathrm{F814} > 23$\,mag.  For these objects even binning
of 100 spectral layers of the datacube, does not result in a S/N-ratio
sufficient for reliable 2D Gaussian fits.  However, for stars brighter
than $\mathrm{m}_\mathrm{F814} \lesssim 22$\,mag such fits do
converge.  For 13 of the 24 pointings this direct point-source fit
could be employed.  To get the FWHM values in the binned layers we
used the PAMPELMUSE software (\citealt{kamann2013phd},
\citealt{Kamann2013}).  For the remaining fields we devised a
minimisation scheme utilising compact galaxies.  For this we visually
selected compact galaxies within the FoV of a pointing from the
CANDELS F814W image.  We then convolved these galaxies with 2D
Gaussians of different FWHMs and resampled these convolved images to
the spatial resolution of MUSE.  In a series of $\sim 100 - 200$
binned layers we then determined the $\chi^2$ of the differences
between the convolved and resampled images to the real data.  Finally,
we fitted the sequence of FWHM values at different
wavelengths with a linear function to obtain $p_0$ and $p_1$.  For the
13 pointings that have a star in the field we found that the two
methods agree within 10\% on the derived $\mathrm{FWHM}(\lambda)$
dependence.  This is shown for one example in
Figure~\ref{fig:psffitsjosie}.  We list in
Table~\ref{tab:psffwhmmusewide} all $p_0$ and $p_1$ coefficients that
we used as input for the spatial filtering procedure in
\texttt{LSDCat}.  For pointings with both the stellar- and compact
object-based estimates of the PSF available, we used the coefficients
resulting in $\mathrm{FWHM}(7000\mathrm{\AA})$ being closest to the
autoguider seeing value given in Table~\ref{tab:musewide_log}, which
we consider to be the best seeing estimate at this wavelength.

For the spectral convolution \texttt{LSDCat} uses a 1D Gaussian
template.  Its width needs to be specified as
the velocity FWHM -- $v_\mathrm{FWHM}$ -- in km\,s$^{-1}$.  We fixed
this parameter in our emission line search to
$v_\mathrm{FWHM}=250$\,km\,s$^{-1}$.  We found this to be the best
single value for achieving the highest S/N for the majority of LAEs in
MUSE datacubes (HW17).  Moreover,
taking the instrumental resolution of MUSE into account, this value is
also consistent with the expectations from the distributions of LAE
FWHMs in the literature \citep[e.g.,][]{Dawson2007,Mallery2012,Yamada2012}.
Although the width of the spectral filter is optimised for LAEs, we
emphasise that the resulting S/N of an Gaussian emission line
decreases only as $\sqrt{2k/(k^2+1)}$ if instead of the optimal
$v_\mathrm{FWHM,\,correct}$ a different $v_\mathrm{FWHM,\,incorrect}$
is chosen, where
$  k = v_\mathrm{FWHM,\,incorrect} / v_\mathrm{FWHM,\,correct} $
(HW17; see also \citealt{Zackay2015}).
Moreover, for the same reason the resulting S/N is also quite robust against
moderate shape mismatches between the Gaussian template and the real
emission line profiles.

Equipped with a set of carefully vetted parameters for the
cross-correlation with the matched filter, we then applied the
relevant \texttt{LSDCat} routines \texttt{lsd\_cat\_spatial.py} and
\texttt{lsd\_cat\_spectral.py} to our 24 datacubes. These routines
also propagate our empirically estimated variances accordingly.  As a
result, we obtained 24 new datacubes (S/N-cubes) that contained in
each voxel the detection significance of an emission line being
present at this position in terms of S/N.

\subsubsection{Thresholding and cataloguing of emission line source
  candidates}
\label{sec:thresh-catal-emiss}

\texttt{LSDCat} collects emission line candidates by thresholding the
S/N-cubes from the matched-filtering procedure.  This task is
performed by the routine \texttt{lsd\_cat.py}, which collects all
detections in the form of a catalogue containing their peak S/N value,
the 3D coordinate of the peak, and the numbers of voxels constituting
the detection cluster.

In theory, i.e. for perfect data without artifacts, the voxel values
of the matched-filter processed cubes are directly related to the
probability of rejecting the null hypothesis of no source signal being
present at a voxel position.  Under this assumption the choice of the
S/N threshold can directly be related to the number of expected false
detections in a datacube from its total number of voxels.  In practice,
however, we have to face the difficulties of possible unknown
systematics in the data and the limitations of our empirical noise
estimate (Sect.~\ref{sec:effective-noise}).  Moreover, as we describe
in Sect.~\ref{sec:classification}, our source classification scheme is
semi-automatic and requires a careful visual inspection of most
sources.  This necessitates a low ratio of spurious to real
detections.

By successively lowering the detection threshold and visual inspection
of the detected sources, we found that the number of unclassifiable
(likely spurious) sources increased rapidly when we lowered the S/N
threshold below eight.  This value represents \textit{``the point of
  diminishing returns''} that we adopted for our emission line search.

\subsection{Classification and cleaning}
\label{sec:classification}

\begin{figure*}[tb]
  \centering
  \includegraphics[width=0.99\textwidth]{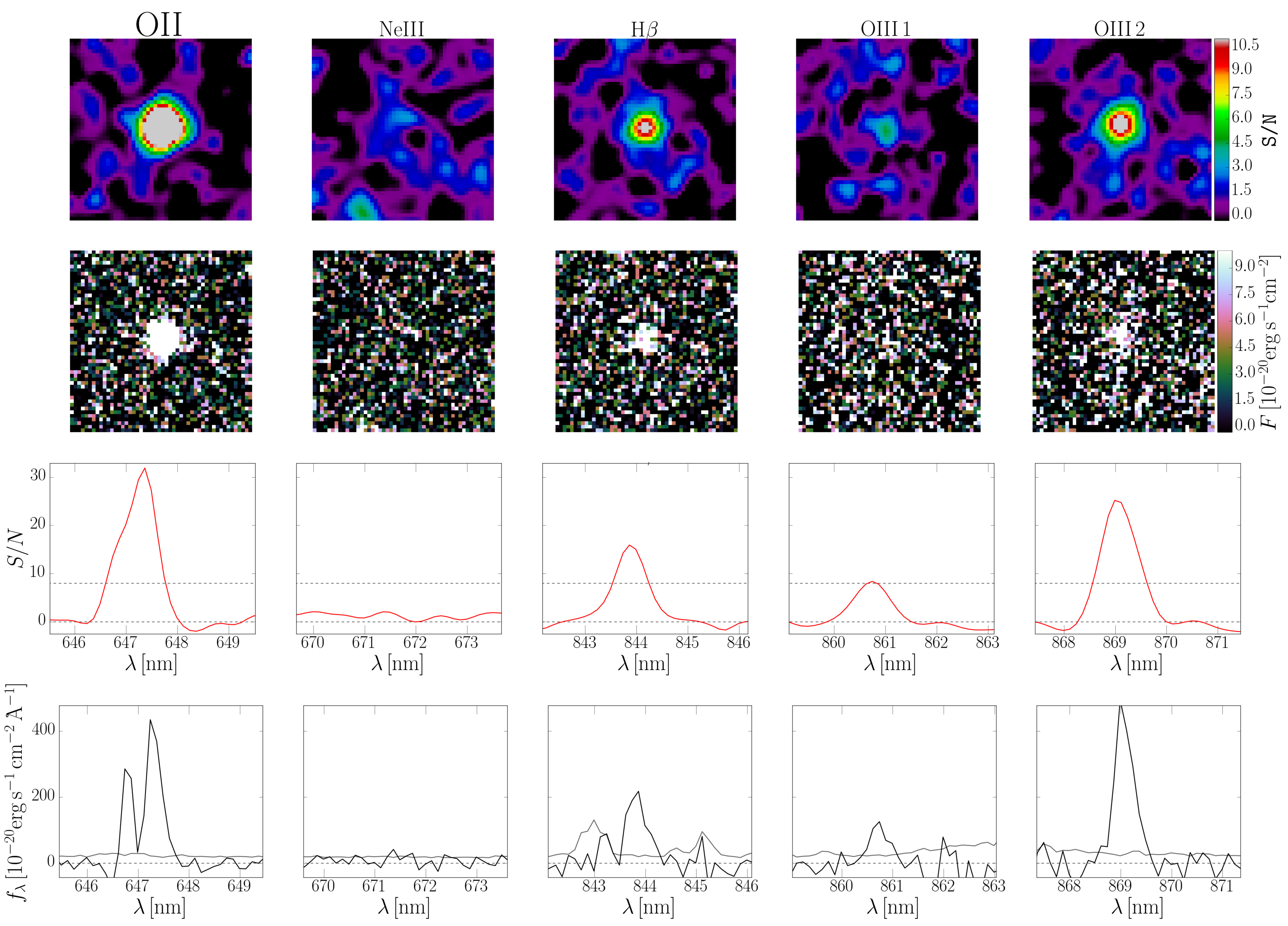}
  \caption{Example of a quality A (confidence 3) object (MUSE-Wide ID 108025145 at
    $z=0.74$, strongest line in S/N is \ion{O}{ii}).  Multiple
    emission lines from this galaxy are detected above the detection
    threshold $\mathrm{S/N}_\mathrm{thresh}=8$. \emph{First row}:
    Individual layers from the S/N cube after matched filtering with
    \texttt{LSDCat}.  The position of the layers is chosen to match
    the classified redshift of the object. \emph{Second row}: Pseudo
    narrow-band images created by integration over 5\,\AA{} around the
    expected position of the emission lines in the continuum
    subtracted datacube. \emph{Third row:} Corresponding segments of
    the spaxel from the \texttt{LSDCat} generated S/N cube with the
    highest S/N peak. The dotted line indicates the detection
    threshold $\mathrm{S/N}_\mathrm{thresh}=8$.  \emph{Fourth row:}
    Corresponding segments of an aperture extracted spectrum (aperture
    radius = 0.7\arcsec{} = 3.5 spectral pixels) from the flux
    datacube.  The grey line shows our empirically determined noise
    estimate.  }
  \label{fig:qualia}
\end{figure*}
\begin{figure*}
  \centering
  \includegraphics[width=0.99\textwidth]{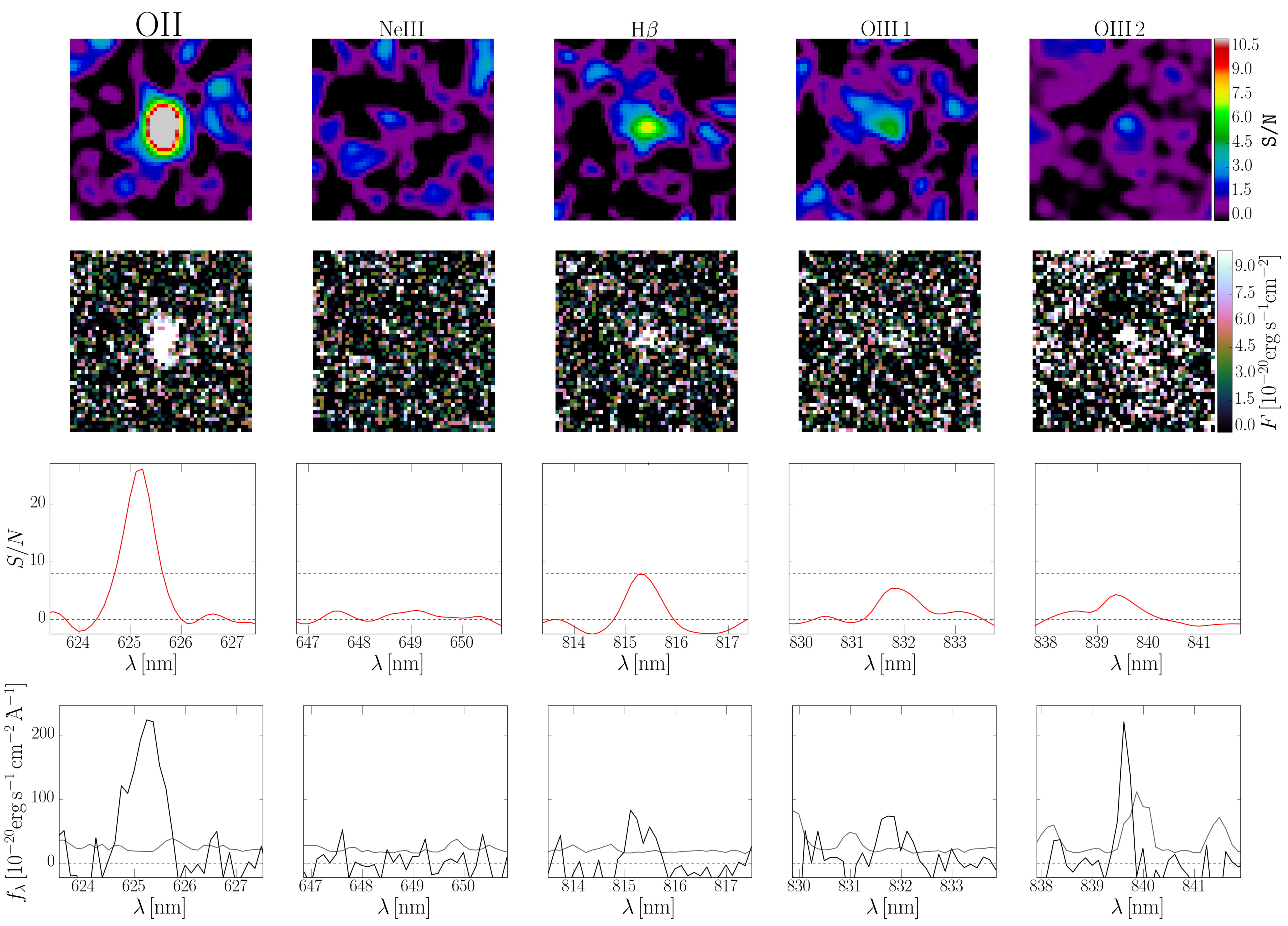}
  \caption{Similar to Fig.~\ref{fig:qualia}, but for a quality B
    (confidence 3) object (MUSE-Wide ID 107021114, strongest line in
    S/N is \ion{O}{ii}).  Only one emission line is detected above the
    detection threshold, but other lines are clearly visible in the
    \texttt{LSDCat} generated S/N cube and in the flux datacube. In
    this object the [\ion{O}{III}] $\lambda 5007$ line falls on a sky
    emission line (as can bee seen by the increased noise level in the
    bottom right panel), so its S/N ratio is lower compared to the
    intrinsically weaker [\ion{O}{III}] $\lambda 4959$ line.}
  \label{fig:qualib}
\end{figure*}
\begin{figure*}
  \centering
  \includegraphics[width=0.88\textwidth,trim=0 0 50em 0,clip]{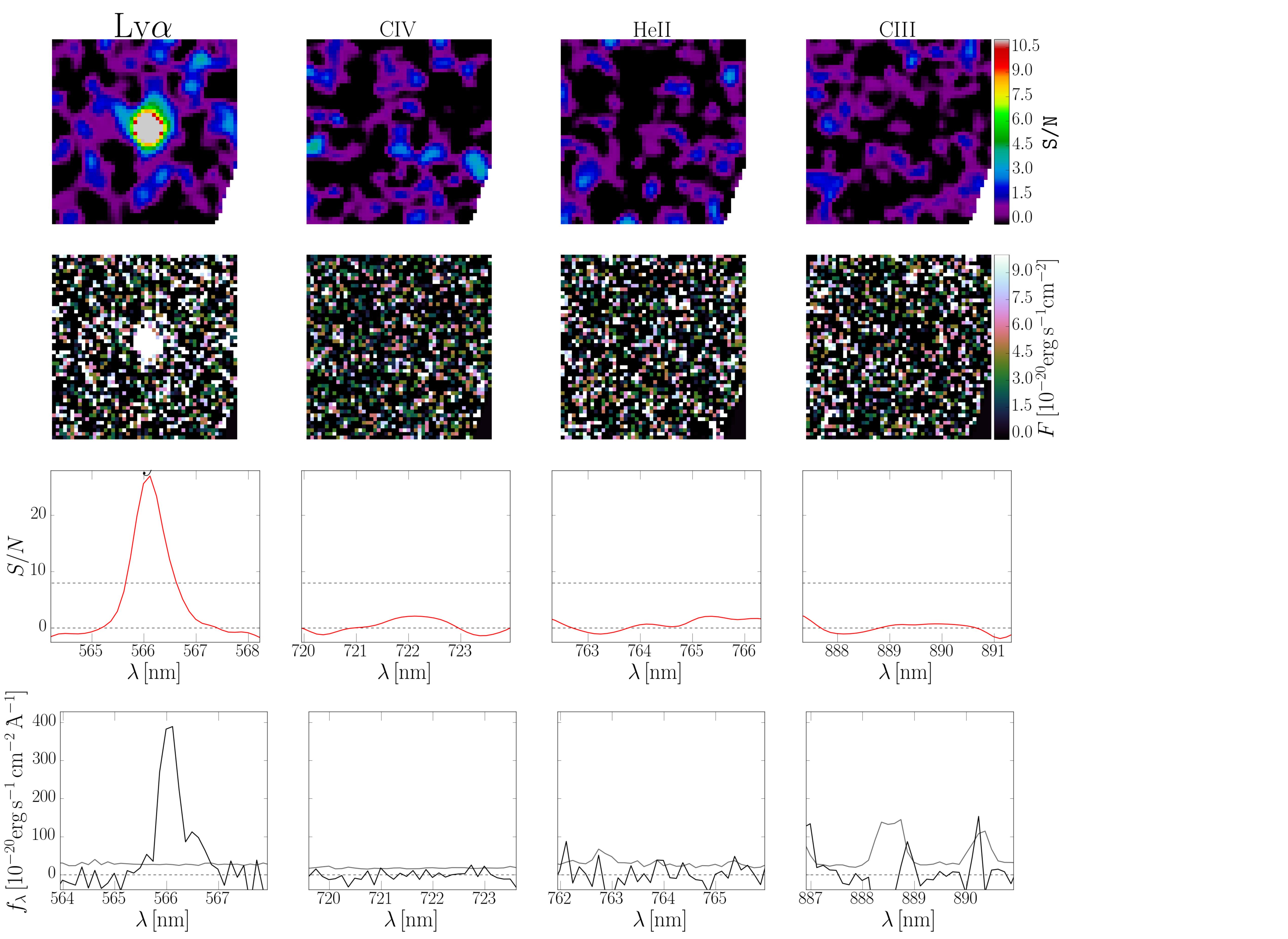}
  \caption{Similar to Fig.~\ref{fig:qualia}, but for a quality C
    (confidence 3) object (MUSE-Wide ID 104015052, only one detected
    line). No other lines were found in the datacube. Also no veto
    lines were found if we would assume the detected line is an
    [\ion{O}{ii}] emission line. Based on the characteristic profile
    of the emission line we confidently classified it as Ly$\alpha$.}
  \label{fig:qualic}
\end{figure*}
\begin{figure*}
  \centering
  \includegraphics[width=0.8\textwidth]{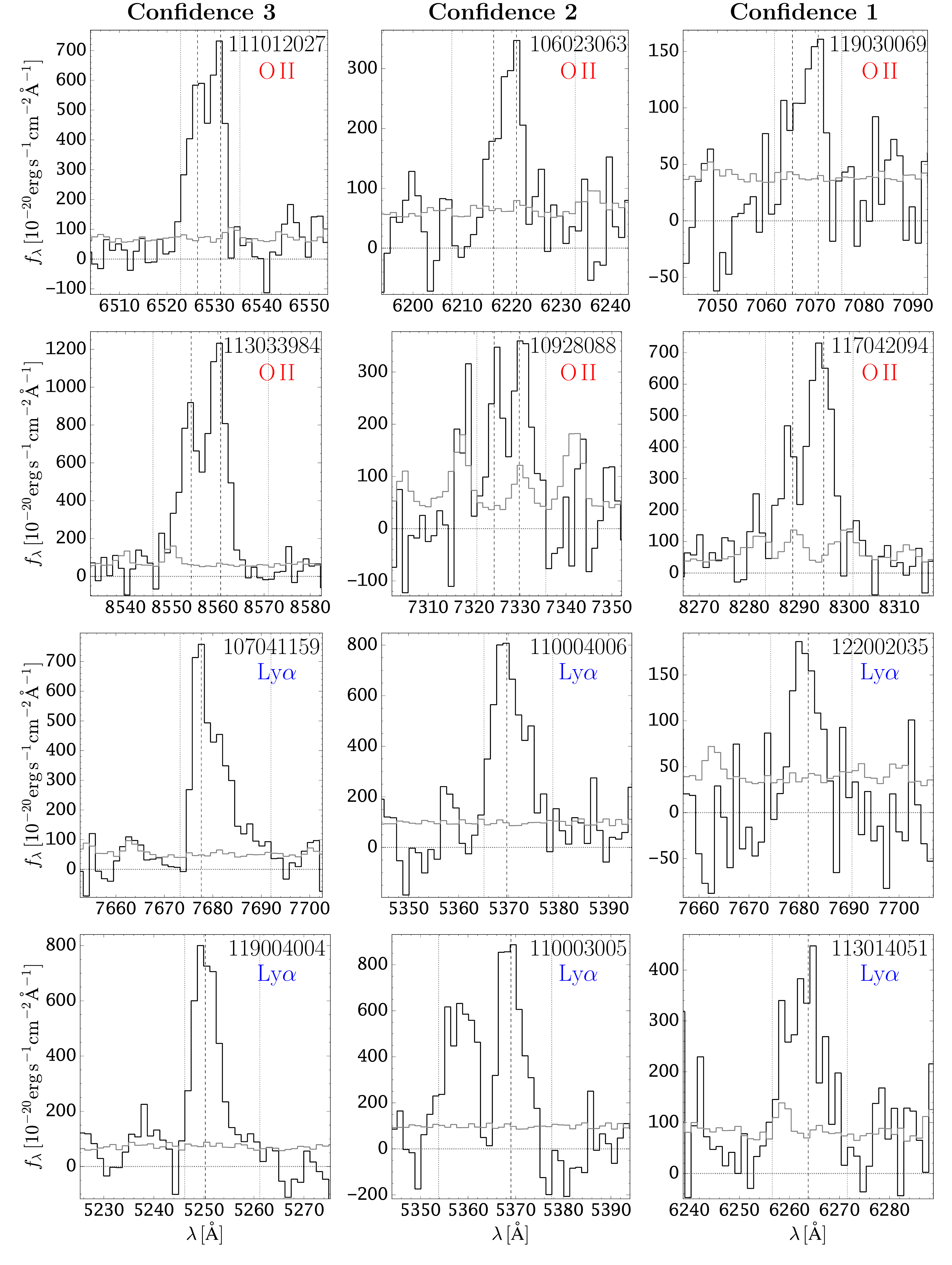}
  \caption{Six representative [\ion{O}{ii}] and Ly$\alpha$ emission
    line profiles from our sample.  The first, second, and third
    column include objects with confidence flag 3, 2, and 1,
    respectively.  In each panel we show extracted spectra from the
    datacube in black and their corresponding variances in grey.  The
    dashed vertical lines mark the observed wavelengths of the
    [\ion{O}{ii}] or Ly$\alpha$ emission according to our determined
    redshift for that object (see Sect.~\ref{sec:redshift}).  The
    dotted vertical lines mark the window used for flux integration by
    LSDCat (see Sect.~\ref{sec:line-source-param}).  In each panel the
    object's MUSE-Wide ID (cf. Sect.~\ref{sec:object-table}) is
    indicated in the top right corner.}
  \label{fig:mw_lya_ex}
\end{figure*}

With the S/N threshold of eight \texttt{LSDCat} provided us with a
catalogue of 2603 line detections over all 24 fields.
% We describe how we classified these line detections into sources below
% in Sect.~\ref{sec:classification}.  
The process of identifying those individual detections, grouping
them into to objects, as well as purging spurious detections from the
initial catalogue was the obvious next step.

\texttt{LSDCat} groups multiple line detections together if they are
within a search radius of 0.8\arcsec on the sky.  Our initial
catalogue contained 374 groups consisting of two or more detections
and 642 spatially isolated single detections.  We inspected those
emission line groups and single line detections with an interactive
graphical tool QtClassify developed especially for this task.  The
functionality and appearance of QtClassify are detailed in Appendix
\ref{sec:qtclassify}.  With this tool all groups and single line
detections were inspected independently by three investigators (ECH,
JK, and TU).  Afterwards these individual classifications were
consolidated into a final classification for each object.  During
consolidation a final quality and confidence value were assigned to
each detection.  While our quality value indicates the amount of
objective information within the datacube and the cross-correlated S/N
datacube that aided the decision process for a particular
classification, our confidence value is a more subjective measure of
``belief'' towards the final classification.

In detail we assign quality flags A, B, and C according to the
following criteria:
\begin{itemize}
\item \emph{Quality A}: Multiple emission lines were detected at the
  same location on the sky and it was possible to anchor a unique
  redshift solution for the object.  These unambiguous identifications
  were assigned automatically in QtClassify but were confirmed by
  visual cross-checks.  We show an example of a quality A classified
  object in Fig.~\ref{fig:qualia}.  There are 288 ``Quality A''
  objects in the final catalogue (34.7\%).
\item \emph{Quality B}: Only one emission line was detected above the
  detection threshold.  However, one or more other lines below the
  detection threshold could also be seen.  These, mostly unambiguous,
  identifications were assigned manually in QtClassify.
  We show an example of a quality B classified object in
  Fig.~\ref{fig:qualib}.  There are 117 ``Quality B'' objects in the final
  catalogue (14\%).
\item \emph{Quality C}: Only one emission line was detected and no
  secondary lines are visible.  The identification was based on the
  visual appearance of the line (e.g. double peak matches
  [\ion{O}{II}] $\lambda\lambda$3726,3728 profile, characteristic
  Ly$\alpha$ profile shape), and the appearance of the object in the
  HST CANDELS images.  Prior information on the photometric redshifts
  was not included.  We show an example of a quality C classification
  in Fig.~\ref{fig:qualic}.  There are 426 ``Quality C'' objects in
  the final catalogue (51.3\%).  For these objects the confidence
  measure indicates how sure the investigators are on a certain
  classification.
\end{itemize}

\noindent
The confidence values 3, 2, and 1 indicate the following:
\begin{itemize}
\item \emph{Confidence 3}: We are certain of the classification.  All
  quality A and most quality B identifications have confidence 3.  For
  single line detections these objects are characterised by a very
  well resolved [\ion{O}{ii}] double-peak or a characteristic
  Ly$\alpha$ profile.  In total we have 578 objects marked with a
  confidence value 3 in the final catalogue (70\%).
\item \emph{Confidence 2}: We are quite sure of the classification,
  but not with the same degree of certainty as for confidence 3. Often
  the reason was simply a somewhat lower S/N of the detection, or a
  remaining (however slight) possibility that another line might
  mimique the appearance of the classified line.  Such cases were
  discussed among the classifiers in the consolidation process.  Often
  the line profile had to be evaluated in detail and examined in
  apertures of different sizes in QtClassify.  In some cases
  also CANDELS images and photometry were inspected in the
  consolidation process.  However, we did not consult any photometric
  redshift catalogues.  210 objects in the final catalogue are marked with
  confidence 2 (25.3\%).
\item \emph{Confidence 1}: We are unsure regarding the classification
  but are certain that the detection is not spurious.  Often the
  classifiers initially disagreed on the classification.  Inspection
  of the line profile and HST imaging data did not resolve the
  uncertainty.  The final classification represents our ``best
  guess''. These lines usually show the lowest S/N-ratios.  Only 43
  objects, i.e. 5.2\% of all objects in our catalogue, are in this
  category.
\end{itemize}
To illustrate our confidence measure we show several line profiles of
emission lines that were classified either as Ly$\alpha$ or
[\ion{O}{ii}] in Fig.~\ref{fig:mw_lya_ex}.

During the consolidation session we also flagged spurious detections
(e.g., telluric line residuals, or detections caused by increased
noise near the FoV borders) and detections caused by residuals due to
imperfect continuum subtraction.  Only 74 of the 2603 line detections
in the initial catalogue were flagged as spurious.  Moreover 636 line
detections were caused by continuum residuals, most of them were
bundled up in very few objects, cold stars or bright early-type
galaxies for which the median filter subtraction does a poor job of
removing the continua, with 30 or more detections each.  These
detections were removed from the final emission line table.  Moreover,
due to the overlap of the pointings, some sources were detected twice
in adjacent pointings.  In the final catalogue we tabulate for such
sources only the quantities determined for the detections in the
pointing where the source is located farthest away from the edge,
where the measurements are less affected by possible edge effects.
Furthermore, some low-$z$ galaxies tended to fragment into detections
of, e.g., multiple \ion{H}{ii} regions.  Such fragmented detections
were manually merged into single objects.  The removal of
double-detections and manual cleaning of fragmented objects
removed  241 emission line detections from the initial
catalogue.  Finally, it turned out that eight of the inspected line
groups (i.e. multiple detections within a 0.8\arcsec{} radius) are not
one object, but are a by chance superposition of two objects at
different redshifts.

By construction, all objects with quality flag \emph{A} have a
confidence value of 3 in the final catalogue.  Most of the objects
with quality \emph{B} also have a confidence value of 3 (107), but for
a few objects (10) the S/N of the additional lines was very low so
that we assigned them with a confidence value of 2.  None of the
quality \emph{B} objects has a confidence value 1. In the quality
\emph{C} class 183 of the 426 single line detections got assigned a
confidence value of 3, 200 of them got assigned a confidence value of
2, and for 43 objects we were unsure regarding the final
classification (confidence value 1).

After the consolidation and cleaning steps of the initial catalogues
described above we arrive at a final catalogue of 1652 emission lines
from 831 emission line galaxies.

\subsection{Positions, spatial extents, and flux measurements}
\label{sec:line-source-param}

\texttt{LSDCat} determines the positions, spatial extents, and fluxes
of detected emission lines with the routine
\texttt{lsd\_cat\_measure.py}.  As source positions in our catalogue
we primarily use the first central moments that are determined in a
pseudo narrow-band image, generated from summing over several layers
in the matched-filtered version of the datacube.  The layers used in
this summation are given by the spectral coordinates of voxels that
are above a certain analysis threshold $\mathrm{S/N}_\mathrm{ana}$ in
the S/N-cube.  After visual inspection of the line profiles, we found
that setting $\mathrm{S/N}_\mathrm{ana} = 3$ delivers a band that is
optimally suited for almost all emission lines.  However, currently
\texttt{LSDCat} does not offer an algorithm to deblend close-by
neighbouring sources.  For those sources the first central moments can
be ambiguous.  In such cases we tabulate as primary coordinate the S/N
peak position introduced in Sect.~\ref{sec:thresh-catal-emiss}.  These
cases are identified by searching in the \texttt{LSDCat} output
catalogues for detections where the S/N-peak position differs
significantly from the first-moment coordinate ($\geq$0.5\arcsec{};
cf. Sect.~\ref{sec:comparision-with-3d}, where we comment on the
astrometric precision of the catalogue).  More details on the
available coordinates per emission line are given in
Sect.~\ref{sec:source-catalogue} where we describe the contents of the
final source catalogue.

\texttt{LSDCat} also measures the spatial extents of our detections by
calculating the characteristic light distribution weighted radius
introduced by \cite{Kron1980}.  To calculate the Kron radius
$R_\mathrm{Kron}$ the \texttt{LSDCat} algorithm follows closely the
\texttt{SExtractor} implementation \citep{Bertin1996}.  The
calculation is performed on the same pseudo narrow-band images that
were used for the determination of the centroids above.  Since for
low-S/N detections the Kron radius can become erroneously small, we
limit the boundary to $R_\mathrm{Kron}^\mathrm{min}=0.6$\arcsec{} in
such cases.  This ensures that the smallest aperture diameter for the
flux measurement, described below, is always larger than the FWHM of
the seeing disk.

Finally, we used \texttt{LSDCat} to measure the fluxes of all emission
lines.  To do this, first, the algorithm creates pure line emission
images by summing up layers containing only the emission line signal.
As above, the bandwidth of these images is given by the spectral
coordinates of the voxels above $\mathrm{S/N}_\mathrm{ana} = 3$ in the
S/N-cube.  We then integrate the flux in these images within
$k\times R_\mathrm{Kron}$ apertures, with $k=1$, $k=2$, $k=3$, and
$k=4$.  The $k=3$ aperture is expected to contain $>95\%$ of the total
flux for compact sources whose light-profile is mainly determined by
PSF broadening \citep[e.g.][]{Graham2005}.  Moreover, we show in the
\texttt{LSDCat} publication (HW17) that automatically determined
fluxes based on the \texttt{LSDCat} $k=3$ aperture compare well with
manually determined fluxes based on a curve-of-growth method.

We caution that for some double peaked Ly$\alpha$ emission lines in
our catalogue we find that the spectral width determined by
\texttt{LSDCat} does not always encompass the weaker bluer peak of the
profile in its entirety (e.g., object 110003005 in
Fig.~\ref{fig:mw_lya_ex}).  In a few cases it even misses the blue
peak completely (e.g., objects 107041159, 11004006, and 119004004 in
Fig.~\ref{fig:mw_lya_ex}).  In particular, 87 of our 237 Ly$\alpha$
emitters show a blue peak and in 53 of those the blue peak is not or
only partially included in the automatically determined flux
integration bandwidth.  Fitting the blue and red peak simultaneously
in the 1D extracted spectra (described in Sect.~\ref{sec:1d-spectra}),
we find that the average flux loss for those 53 objects is 10\%.
Rather than manually changing the spectral integration width for those
sources we opt for providing a homogenised set of automatically
determined flux measurements.  We will address the accurate
measurement of LAE fluxes in a forthcoming publication.  Moreover, the
fluxes in automatically determined $3\,R_\mathrm{Kron}$ apertures are
also not robust for galaxies that are exceptionally extended or have
close by companions.  For such objects emission line flux ratios will
likely be distorted.  Another caveat of the provided fluxes is, that
for all except eight [\ion{O}{ii}] detections the
$\lambda\lambda$3726,2729\AA{} doublet is detected as a single line,
thus the tabulated flux is integrated over both lines.  We encourage
users interested in more accurate emission line flux measurements to
exploit the information contained in the 3D source datacubes that we
provide with this catalogue (Sect.~\ref{sec:source-datacubes}).

\subsection{Redshift measurements}
\label{sec:redshift}

The line detection and identification gives us an approximate measure
of the redshift which we improved as described in the following.  To
accurately measure redshifts of the 831 emission line galaxies in our
sample we use 1D spectra that we extracted for each of our objects.
These spectra are released with the catalogue and the extraction is
described in Sect.~\ref{sec:1d-spectra}.  Depending on whether the
source is a high-$z$ LAE or a low-$z$ galaxy detected by its
rest-frame optical emission line we employed different emission line
fitting strategies.  We describe the fitting method in the following
two subsections.  All our redshifts are vacuum redshifts within the
barycentric reference frame.

% In Sect.~\ref{sec:comparision-with-3d} we compare
% our redshifts with the compilation of redshifts in the CANDELS/Deep
% region of the GOODS-South given by \cite{Momcheva2016} and with the
% redshifts from the first data release of the VIMOS Ultra Deep Survey
% \citep{LeFevre2015,Tasca2016}.

\subsubsection{Determining Lyman $\alpha$ galaxy redshifts}
\label{sec:high-redshift-lyman}

To determine the redshifts of the LAEs we fitted the Ly$\alpha$ line
profiles with the formula
\begin{equation}
  \label{eq:asymeq}
  f(\lambda) = A \times \exp \left \{ - 
     \frac{(\lambda - \lambda_0)^2}{2 \times (a_\mathrm{asym}(\lambda -
      \lambda_0) + d)^2} \right \} 
 %\; \text{.}
\end{equation}
introduced by \cite{Shibuya2014}.  Eq.~(\ref{eq:asymeq}) describes an
asymmetric Gaussian profile used to fit Ly$\alpha$ profiles.
\cite{Shibuya2014} argue that Eq.~(\ref{eq:asymeq}) provides a more
robust peak wavelength for the typical LAE profiles than a simple
Gaussian.  The free parameters $A$, $\lambda_0$, $a_\mathrm{asym}$,
and $d$ in our fit to Eq.~(\ref{eq:asymeq}) are the amplitude, the
peak wavelength, the asymmetry parameter, and the typical width of the
line, respectively.  It is known and commonly attributed to the
complex Ly$\alpha$ radiative transfer physics that Ly$\alpha$ redshift
measurements are systematically offset by
$\sim100$\,--\,$200$\,km\,s$^{-1}$ with respect to the systemic
redshift determined from rest-frame optical emission lines
\citep[e.g.][]{McLinden2011,Rakic2011,Chonis2013,Erb2014,Song2014,Hashimoto2015,Trainor2015}. No
correction for such offsets was applied here.

In practical terms, the fitting was performed in a window around the
peak with flux values being greater than 10\,\% of the peak flux.  For the
87 Ly$\alpha$ emitters in our sample that show a double peaked profile
we restricted the fit to the region of the red peak.  For 11 objects
we had to adjust the window size manually to avoid strong
sky-subtraction residuals.
% \footnote{The MUSE-Wide IDs of the objects requiring manual
% intervention are 10105016, 10130050, 10924083, 11101001, 11504089,
% 11018062, 11705019, 11902002, 12104014, 12212068, and 12403012.}.
We accounted for a possible continuum by subtracting a running median
from the 1D spectrum. 
%  The $\lambda_0$ values from the fit to air
% wavelengths are converted to vacuum wavelengths before being
% transformed to redshifts using the laboratory wavelength of Ly$\alpha$
% (Table~\ref{tab:emlines}).
  Finally, we determined the error on each
redshift by repeating the fitting procedure 100 times on random
realisations of the spectra generated by perturbing each spectral
pixel according to the noise statistics of that pixel from the
associated error spectrum.

\subsubsection{$z\lesssim1.5$ galaxies}
\label{sec:1.5-galaxies}

Emission line galaxies at $z\lesssim1.5$ are detected in MUSE
datacubes by the typical strong rest-frame optical emission lines of
star-forming galaxies \citep[e.g.][]{Kennicutt1992}: [\ion{O}{ii}]
$\lambda\lambda$3276,3278 (detected in 472 objects), [\ion{O}{iii}]
$\lambda\lambda 4958, 5006$ (detected in 310 objects), H$\beta$
(detected in 184 objects), and H$\alpha$ $\lambda6563$ (detected in 73
objects).  In Table~\ref{tab:emlines} we list the air- and vacuum
wavelengths of these transitions.  

To determine redshifts we fitted 1D Gaussian profiles to those
emission lines.  The [\ion{O}{ii}] $\lambda\lambda$3276,3278 doublet
was fitted by a double component Gaussian with fixed separation and
free intensity ratio, while all other lines were fitted with single
components.  The continuum was subtracted with a 151 pixel wide
running median in spectral direction.  For objects having several
detected emission lines we computed the S/N-weighted mean redshift of
all emission line fits.  The error on the redshift was determined by
repeating the fitting procedure 100 times on realisations of the
spectra generated by perturbing each pixel according to the noise
statistics from the error spectrum.

\section{Source catalogue, spectra, and  datacubes}
\label{sec:source-catalogue}

With this publication we provide the following data products:
\begin{itemize}
\item A catalogue of all 831 detected emission line objects.
\item A table of all 1652 detected emission lines in those objects.
\item 1D PSF-weighted extracted spectra of the emission line objects .
\item 3D datacubes of the emission line objects.
\end{itemize}
The tabular data is available in its entirety in the electronic
version of the journal.  Moreover, these data is also available via
the MUSE-Science website\footnote{\url{http://muse-vlt.eu/science/muse-wide-survey}} and via the CDS\footnote{Placeholder for CDS
  URL, when assigned}, where also the 1D spectra and 3D datacubes are stored.
% TODO: insert CDS link as soon as we have it.
In the following subsections we describe these data products in
detail.

\subsection{Object Table}
\label{sec:object-table}

\begin{table}
  %\centering
  \caption{Columns of the object table}
  \label{tab:obtab}
  \begin{tabular}{lll} \hline \hline
    Column Name          &  Short Description                  &  Example Entry\tablefootmark{a}   \\ \hline
    \texttt{UNIQUE\_ID}  & Unique MUSE-Wide object ID          &  101001006                                \\
    \texttt{RA}  & Right ascension $\alpha_\mathrm{J2000}$ [deg]  & \phantom{-}53.060185\dots                 \\
    \texttt{DEC} & Declination $\delta_\mathrm{J2000}$ [deg]      & -27.813464\dots                           \\
    \texttt{Z}   & Redshift                            & 0.310564\dots                             \\
    \texttt{Z\_ERR} & Error on the redshift            & 0.000018\dots                             \\
    \texttt{LEAD\_LINE} & Highest S/N detected line\tablefootmark{b}    & \texttt{Ha}              \\ 
    \texttt{SN}  & S/N of the \texttt{LEAD\_LINE}      & 76.106950\dots                            \\
    \texttt{QUALITY} & Quality flag (Sect.~\ref{sec:classification}) & a                           \\
    \texttt{CONFIDENCE} & Confidence value (Sect.~\ref{sec:classification}) & 3                    \\
    \texttt{OTHER\_LINES} & Other detected lines\tablefootmark{b}      & \texttt{O2,Hg,Hb}\dots \\
    \texttt{GUO\_ID} & Associated source in & {} \\
    {} & \cite{Guo2013} catalogue                               & 10720 \\
    \texttt{GUO\_SEP} & Angular separation to \\
    {} & \cite{Guo2013} source [\arcsec{}]   & 0.37\\
    \texttt{SKELTON\_ID} & Associated source in & {} \\
    {} & \cite{Skelton2014} catalogue   & 20736 \\
    \texttt{SKELTON\_SEP} & Angular separation to \\
    {} & \cite{Skelton2014} source [\arcsec{}]  & 0.25 \\
    \hline
  \end{tabular}
  \tablefoottext{a}{Entries containing \dots are truncated compared to the original format of the table.} \tablefoottext{b}{Emission line identifiers are explained in Table~\ref{tab:emlinestab}.}
\end{table}

\begin{table*}
    \centering
    \caption{Columns of the emission line table}
    \label{tab:emlinestab}
    \begin{tabular}{lll} \hline \hline Column & Unit(s) & Description \\
      \hline
      \texttt{UNIQUE\_ID}    & ---  & Unique MUSE-Wide object ID. \\
      \texttt{POINTING\_ID}  & ---  & Pointing Number (see Fig.~\ref{fig:footprint}). \\
      \texttt{OBJ\_ID}       & ---  & Object ID -- only unique per pointing. \\
      \texttt{RID}           & ---  & Running ID -- only unique per pointing. \\
      \texttt{IDENT}         & ---  & Line identification. \\
      \texttt{COMMENT}       & ---  & Free-form comment added during. \\
      {}                     & {}   & classification and cleaning (Sect.~\ref{sec:classification}). \\
      \texttt{SN}            & ---  & Detection significance (Sect.~\ref{sec:cross-corr-with}). \\
      \texttt{\{RA,DEC,LAMBDA\}\_SN} & deg / \AA{}  & 3D S/N-weighted position. \\
      \texttt{\{RA,DEC,LAMBDA\}\_PEAK\_SN} & deg / \AA{}  & S/N-peak position. \\
      \texttt{\{RA,DEC\}\_1MOM} & deg & First central moment coordinate, determined \\
      {}                        &     & in optimal narrow band image (Sect.~\ref{sec:line-source-param}). \\
      \texttt{LAMBDA\_NB\_\{MIN,MAX\}} & \AA{} & Minimum- and maximum wavelength of optimal\\
      {} & {} &   narrow band image.  Used for flux
                integration.\\
      \texttt{R\_KRON} & arcsec & \citeauthor{Kron1980}-Radius (see Sect.~\ref{sec:line-source-param}). \\
      \texttt{F\_KRON,F\_\{2,3,4\}KRON} & 10$^{-20}$erg\,s$^{-1}$\,cm$^{-2}$ & Flux extracted in ($k\times$) \texttt{R\_KRON} aperture within \\
      {}                                & {}                     &  narrow band defined by \texttt{LAMBDA\_NB\_\{MIN,MAX\}}. \\
      \texttt{F\_KRON\_ERR,F\_\{2,3,4\}KRON\_ERR} & 10$^{-20}$erg\,s$^{-1}$\,cm$^{-2}$ & Error on the extracted flux. \\
      \texttt{BORDER\_FLAG}  & --- & Flag indicating whether 3$\times R_\mathrm{Kron}$  overlaps with FoV border. \\
      \hline
    \end{tabular}
\tablefoot{Comma separated list within curly braces in the first column indicate the set of similar columns.  Wavelengths are vacuum wavelengths.}
\end{table*}

\begin{table*}
  \centering
  \caption{Detected emission lines in the catalogue}
  \label{tab:emlines}
    \begin{tabular}{llllcc} \hline \hline
      Transition(s)                                  & \texttt{IDENT}\tablefootmark{a}     & \multicolumn{1}{c}{Air Wavelength(s)\tablefootmark{b}}            & \multicolumn{1}{c}{Vacuum Wavelength(s)\tablefootmark{b}}  & Detections & Comments\tablefootmark{c}  \\
      {}                                           &                    & \multicolumn{1}{c}{[\AA{}]}                   & \multicolumn{1}{c}{[\AA{}]}            & {}         &        \\ \hline
      \ion{O}{vi} $\lambda\lambda$1032,1038        & \texttt{O6\_2}     & {\dots}                   & 1031.93, 1037.62   & 2          & (1)      \\
      Ly$\alpha$  $\lambda$1216                    & \texttt{Lya}       & {\dots}                   & 1215.670           & 242        & (2),(3)    \\ 
      \ion{N}{v}  $\lambda\lambda$1238,1242        & \texttt{N5}        & {\dots}                   & 1238.821, 1242.804 & 1          & (4)      \\ 
      \ion{C}{iv} $\lambda\lambda$1549,1550        & \texttt{C4}        & {\dots}                   & 1548.203, 1550.777 & 3          & (3),(4)    \\ 
      \ion{Mg}{ii} $\lambda$2796                   & \texttt{Mg2}       & 2795.528                  & 2796.290           & 1          & {}     \\ 
      $[$\ion{O}{II}$]$ $\lambda\lambda$3726,3729  & \texttt{O2}        & 3726.032, 3728.815        & 3727.048, 3729.832 & 480        & (3),(5)    \\
      $[$\ion{Ne}{iii}$]$ $\lambda$3869            & \texttt{Ne3}       & 3869.06                   & 3870.115           & 49         & {}     \\
      $[$\ion{Ne}{iii}$]$ $\lambda$3968            & \texttt{Ne32}      & 3967.79                   & 3968.872           & 3          & (6)      \\
      H$\zeta$ $\lambda$3889                       & \texttt{Hzet}      & 3889.049                  & 3890.109           & 2          & {}     \\
      H$\varepsilon$ $\lambda$3970                 & \texttt{Heps}      & 3970.072                  & 3971.154           & 1          & {}     \\
      H$\delta$ $\lambda$4102                      & \texttt{Hd}        & 4101.734                  & 4102.852           & 18         & {}     \\
      H$\gamma$ $\lambda$4340                      & \texttt{Hg}        & 4340.464                  & 4341.647           & 74         & {}     \\
      $[$\ion{O}{iii}$]$ $\lambda$4363             & \texttt{O3\_3}     & 4363.210                  & 4364.400           & 2          & {}     \\
      H$\beta$ $\lambda$4861                       & \texttt{Hb}        & 4861.325                  & 4862.650           & 183        & (3)      \\
      $[$\ion{O}{iii}$]$ $\lambda$4959             & \texttt{O3\_1}     & 4958.911                  & 4960.263           & 135        & (3)      \\
      $[$\ion{O}{iii}$]$ $\lambda$5007             & \texttt{O3\_2}     & 5006.843                  & 5008.208           & 304        & (3)      \\
      \ion{He}{i} $\lambda$5876                    & \texttt{He1}       & 5875.615                  & 5877.217           & 6          & {}     \\
      $[$\ion{O}{i}$]$ $\lambda$6300               & \texttt{O1}        & 6300.304                  & 6302.022           & 3          & {}     \\ 
      $[$\ion{N}{ii}$]$ $\lambda$6549              & \texttt{N2\_1}     & 6548.04                   & 6549.825           & 4          & {}     \\ 
      H$\alpha$ $\lambda$6563                      & \texttt{Ha}        & 6562.80                   & 6564.589           & 73         & (3)      \\
      $[$\ion{N}{ii}$]$ $\lambda$6583              & \texttt{N2\_2}     & 6583.46                   & 6585.255           & 19         & {}     \\
      $[$\ion{S}{ii}$]$ $\lambda$6717              & \texttt{S2\_1}     & 6716.44                   & 6718.271           & 30         & {}     \\
      $[$\ion{S}{ii}$]$ $\lambda$6730              & \texttt{S2\_2}     & 6730.81                   & 6732.645           & 17         & {}     \\
      \hline
    \end{tabular}
    \tablefoot{
      \tablefoottext{a}{Emission line identifier code used in the \texttt{IDENT} column of the emission line table described in Sect. \ref{sec:emissin-line-table}.} \\
      \tablefoottext{b}{Wavelengths are from the ``Atomic Line List'' compiled by
        P. van Hoof: 
        \url{http://www.pa.uky.edu/~peter/atomic/}.
        Wavelengths longward of 2000\,\AA{} were queried in air wavelengths and
        converted to vacuum wavelengths with the formula used in the Vienna atomic line
        database \citep{Ryabchikova2015}:
        \url{http://www.astro.uu.se/valdwiki/Air-to-vacuum\%20conversion}.} \\
      \tablefoottext{c}{(1): doublet in single object resolved into two individual line detections. 
        (2): 237 LAEs, with five having the double peak profile resolved into two individual line detections. 
        (3): used for redshift determination. 
        (4): doublet unresolved, i.e. always detected as single line. 
        (5): in all except eight detections (103035109, 105019069, 109021072, 109023081, 114015092, 115046179, 124009024, and 124040076) the $[$\ion{O}{II}$]$ $\lambda\lambda$3726,3729 doublet is unresolved.
        (6): blend with H$\varepsilon$.
      }}
\end{table*}

\begin{figure*}
  \centering
  \includegraphics[width=0.45\textwidth]{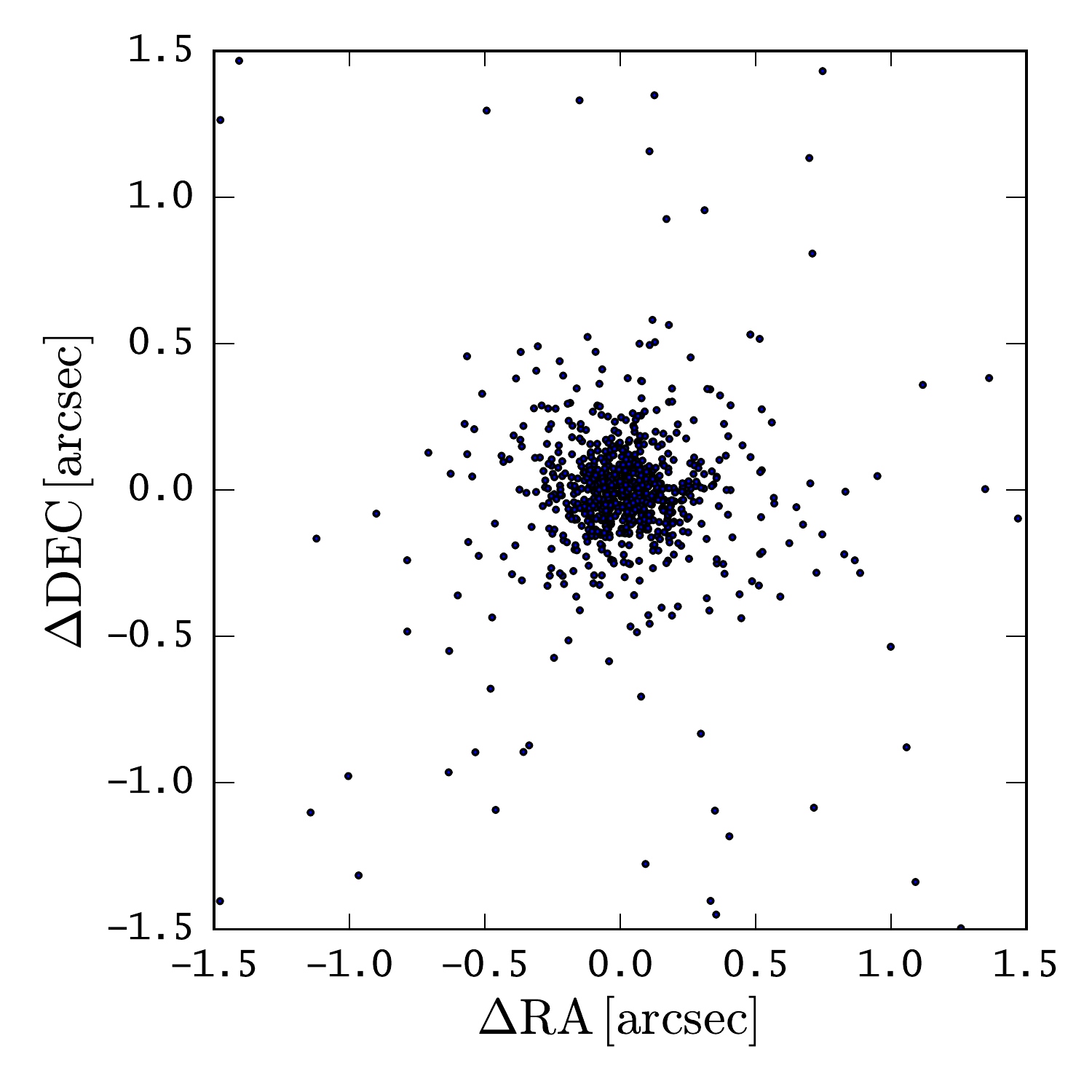}  \includegraphics[width=0.45\textwidth]{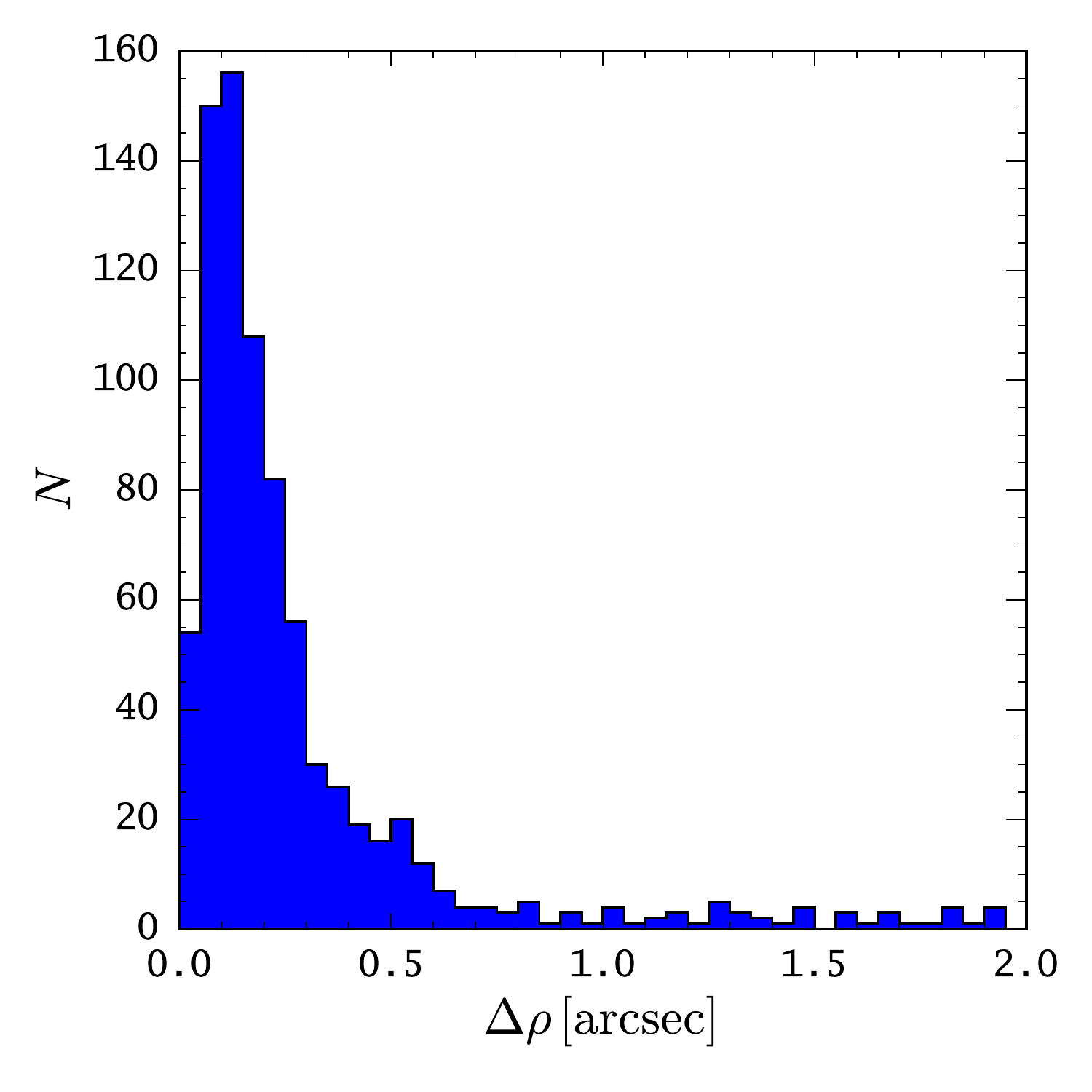}
  \caption{\emph{Left panel}: Relative differences in right ascension
    and declination between object coordinates in the MUSE-Wide
    catalogue and objects with the closest on-sky separation in the
    3D-HST catalogue \citep{Skelton2014}. \emph{Right panel}: Angular
    separation $\Delta \rho$ for between MUSE-Wide objects and most
    closely separated 3D-HST sources.  83\%  of the objects from our
    MUSE-Wide catalogue have cross-matches within 0.5\arcsec{} in the
    \cite{Skelton2014} catalogue.
     % A visual inspection revealed that
 %    catalogue matches with $\Delta \rho > 0.5$\arcsec{} are not
 %    physically associated (see Fig.~\ref{fig:nocounter} for examples).
  }
  \label{fig:deltapos}
\end{figure*}

\begin{figure*}
  \centering
  \includegraphics[width=0.95\textwidth]{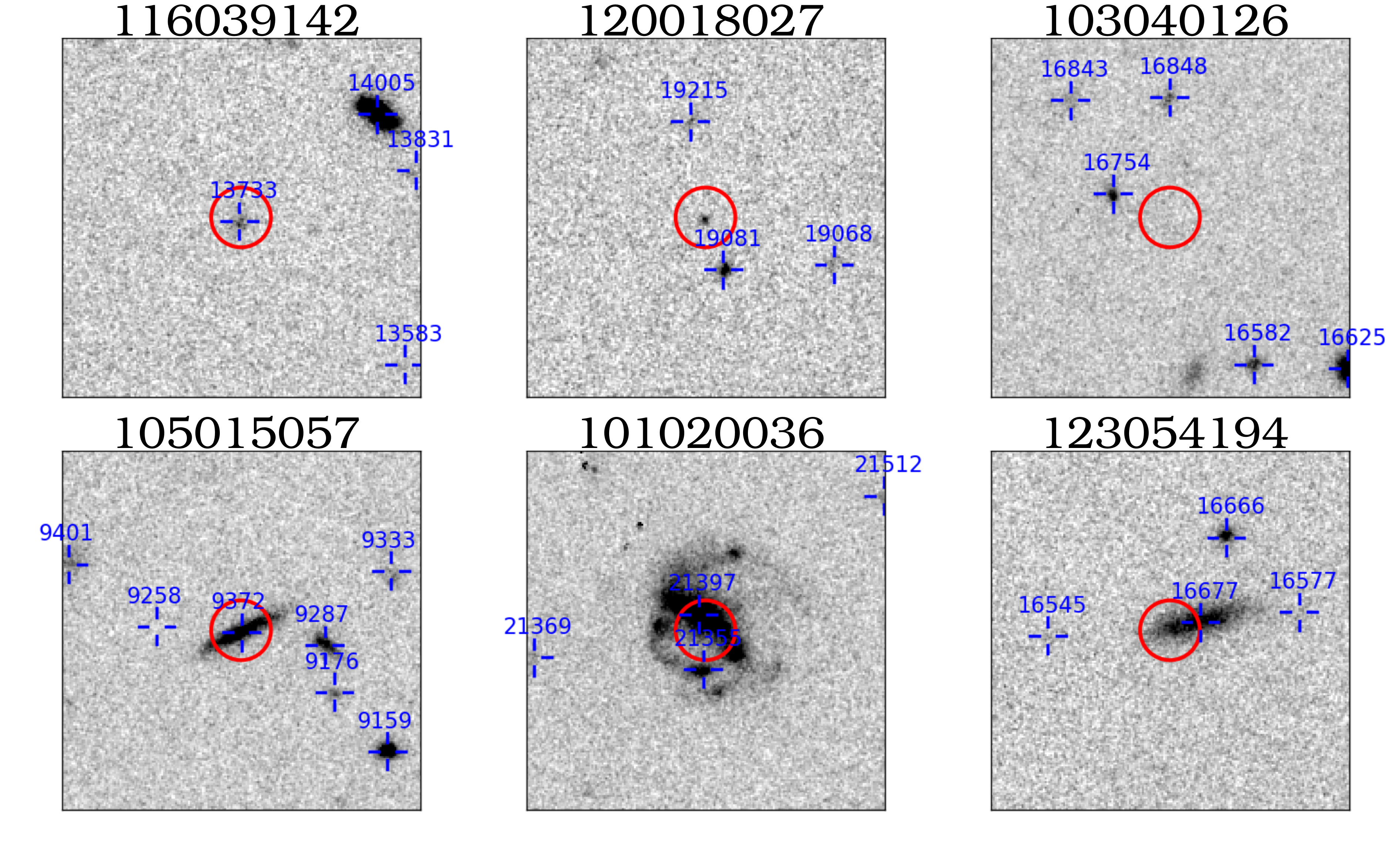}
  \caption{Six example MUSE-Wide emission line sources CANDELS HST
    F814W 6\arcsec{}$\times$ 6\arcsec{}cutouts. The red circle of
    1\arcsec{} diameter is centred on our catalogued position, the
    blue crosses with labels indicate the positions and IDs given in
    the \cite{Skelton2014} 3D-HST photometric catalogue.  The top row
    shows LAEs, one with a catalogued photometric counterpart, one
    with a photometric counterpart and no catalogue entry and one with
    no counterpart in the image.  The bottom row shows three typical
    $z<1.5$ counterparts.}
  \label{fig:nocounter}
\end{figure*}

In Table~\ref{tab:obtab} we present the columns of the catalogue that
contains all 831 detected emission line galaxies in the first 24
pointings (22.2 arcmin$^2$) of the MUSE-Wide survey.  
The details of the 14 columns are given below:
\begin{itemize}
\item  \texttt{UNIQUE\_ID} contains a unique MUSE-Wide
  ID. This ID is composed of nine digits divided into four groups of
  the format \texttt{ABBCCCDDD}.  Here \texttt{A} designates the
  MUSE-Wide survey-area (1$\equiv$ECDF-S CANDELS/Deep, 2$\equiv$COSMOS
  CANDELS (not used in the catalogue described here), or other numbers
  for future MUSE-Wide regions), \texttt{BB} indicates the pointing
  number (here 01--24, see Fig.~\ref{fig:footprint}), \texttt{CCC}
  refers to the per-pointing object ID, and \texttt{DDD} to the
  running ID of the strongest line.  These last two identifiers relate
  to the emission-line table explained in 
  Sect.~\ref{sec:emissin-line-table}.
\item \texttt{RA} and \texttt{DEC} contain the position of the galaxy
  in right-ascension $\alpha_\mathrm{J2000}$ and declination
  $\delta_\mathrm{J2000}$.  For most sources this position is given as
  first central moments determined in an adaptive narrow-band image
  (Sect.~\ref{sec:line-source-param}) of the lead line (see column
  \texttt{LEAD\_LINE} description below).  However, for 40 sources
  this position differed by more than 0.5\arcsec{} to the peak S/N
  position found in the initial thresholding step
  (Sect.~\ref{sec:thresh-catal-emiss}).  Visual inspection revealed
  that these cases are often affected by blends with neighbouring
  sources.  Since the peak S/N-coordinate is less affected by
  blending, for those 40 cases we replace the first central moment
  coordinate with the peak S/N coordinate.  We also provide more
  positional parameters per emission line detection in
  Sect.~\ref{sec:emissin-line-table}.
\item \texttt{Z} contains the redshift $z$ for each galaxy and
  column \texttt{Z\_ERR} contains the error on this quantity as
  explained in Sect.~\ref{sec:redshift}.
\item \texttt{LEAD\_LINE} contains the lead line for each galaxy.  The
  lead line is defined as a galaxy's emission line that has the
  highest S/N after the matched-filtering process
  (Sect.~\ref{sec:cross-corr-with}).  The lead line is therefore not
  necessarily the line with the highest flux.  Potential identifiers
  encountered in the \texttt{LEAD\_LINE} column are tabulated in
  Table~\ref{tab:emlines}.
\item  \texttt{SN} contains the S/N of the lead line.
\item \texttt{QUALITY} and \texttt{CONFIDENCE} tabulate the quality
  and confidence values indicating the robustness of the object
  classification and line identification as described
  Sect.~\ref{sec:classification}.
\item \texttt{OTHER\_LINES} we provide a comma-separated
  list of all other emission lines that were detected above the
  detection threshold, but at a lower S/N than the lead line.  This
  field is empty in case of a single line detection.  Potential
  identifiers encountered in the \texttt{OTHER\_LINES} column are
  tabulated in Table~\ref{tab:emlines} (see also
  Sect.~\ref{sec:emissin-line-table} below).
\end{itemize}

In the last four columns of the object-table we provide cross-matches
(and angular separations) to existing catalogues based on source
detection in HST imaging of the CANDELS-GOODS-S field.
\begin{itemize}
\item \texttt{GUO\_ID} contains the object identifier of the
  cross-matched source from the CANDELS~GOODS-S catalogue
  \citep{Guo2013}. Its value is set to 0 for objects where no
  counterpart could be assigned.
\item  \texttt{GUO\_SEP} gives the angular separation between the
  MUSE-Wide position and \cite{Guo2013} source.
\item \texttt{SKELTON\_ID} contains the in object identifier of the
  cross-matched source from the 3D-HST/CANDELS photometric catalogue
  \citep{Skelton2014}.  It is set to 0 for objects where no
  counterpart could be assigned.
\item \texttt{SKELTON\_SEP} gives the angular separation between
  the MUSE-Wide position and the \cite{Skelton2014} source.
\end{itemize}
These cross-matches where obtained by first selecting for each object
in our catalogue the corresponding object with the smallest on-sky
separation in the photometric catalogues.  In Fig.~\ref{fig:deltapos}
we show the relative distances in right ascension and declination with
respect to the \cite{Skelton2014} catalogue, as well as histogram of
the angular separations $\Delta \rho$.  Both panels of Figure look
almost identical if we compare to the positions from the
\cite{Guo2013} catalogue instead.  The isotropic distribution around
$(\Delta\mathrm{RA}, \Delta\mathrm{Dec}=(0,0)$ indicates that there is
no systematic off-set in the positions reported in our catalogue.  The
median $\Delta \rho$ of this catalogue cross-match is 0.17\arcsec{}
(0.19\arcsec) and 83\% (80\%) of all MUSE-Wide objects have a
photometric counterpart in \cite{Skelton2014} catalogue (the
\citealt{Guo2013} catalogue) within 0.5\arcsec{}.  Visual inspection
of F814W thumbnails with catalogue positions overlaid (see examples in
Fig.~\ref{fig:nocounter}) reveals that all photometric counterparts
with $\Delta \rho \leq 0.5$\arcsec{} are certain associations.  Based
on this visual screening we included 34 and 36 additional counterparts
with angular separations $\geq 0.5$\arcsec{} from the
\cite{Skelton2014} and \cite{Guo2013} catalogue, respectively.  For
these sources either the broad-band photometric centroid appears to be
offset from the emission-line centroid, or in the case of some
low-redshift sources the photometric catalogues tend to separate one
source into multiple detections (see, e.g., ID 101200036 in bottom row
of Fig.~\ref{fig:nocounter}).

\subsection{Emission Line Table}
\label{sec:emissin-line-table}

For each of the 1652 detected emission lines listed in columns
\texttt{LEAD\_LINE} and \texttt{OTHER\_LINES} of the object catalogue
\texttt{LSDCat} outputs a set of measurements.  We supplement the
object catalogue by providing a table that contains these measurements
as well as additional information for all emission lines.  This table
is available only in electronic format at the CDS as a FITS table and
contains the following columns, which are also briefly summarised in
Table~\ref{tab:emlinestab}:
\begin{itemize}
\item \texttt{UNIQUE\_ID} contains the unique identifier for each
  detected galaxy in our survey (cf. Sect.~\ref{sec:object-table}).
  Therefore, this column establishes the link between the object table
  and the emission line table.
\item \texttt{OBJECT\_ID} is a unique integer identifier for each
  identified object within a MUSE-Wide pointing.  This identifier
  separates spatially overlapping objects, i.e. two galaxies at
  different redshifts but at the same position on the sky.  This
  integer number, padded with zero digits on the left, comprise the
  digits \texttt{CCC} of \texttt{UNIQUE\_ID}.  While in the present
  catalogue never more than 100 objects are within a pointing, we
  reserve the third digit for future MUSE-Wide catalogue publications
  to ensure a unified system across different MUSE-Wide catalogues.
\item \texttt{RID} is the running index from our initial
  \texttt{LSDCat} catalogues after thresholding
  (Sect.~\ref{sec:thresh-catal-emiss}) and thereby uniquely indexes
  each detected emission line in a pointing.  This integer number,
  padded with zeros on the left, comprise the digits \texttt{DDD} of
  \texttt{UNIQUE\_ID}.  Since we removed 951 of the detections in the
  classification and cleaning process of the initial catalogue
  (Sect.~\ref{sec:classification}), the \texttt{RID} column is not a
  running integer index per pointing in the final catalogue.
\item \texttt{IDENT} contains the identification of each
  catalogued emission line established in the classification
  process (Sect.~\ref{sec:classification}).  We encode line
  identifications in a string of four alphanumeric characters.  A
  legend of all those emission lines identifiers, their common
  designations and their wavelengths is given in
  Table~\ref{tab:emlines}.  In this table we also list the total
  number of detections of a particular emission line in our catalogue.
\item \texttt{COMMENT}: During the consolidation of our
  classifications with QtClassify we added comments when necessary.
  These are stored in this column.
\item \texttt{BORDER\_FLAG} contains a logical
  flag, indicating whether the 3 Kron-radii extraction aperture for the emission
  line flux measurement overlaps with the FoV borders of the
  particular pointing in which the flux was extracted.
\end{itemize}
The other columns in the emission line table stem directly from
\texttt{LSDCat}:
\begin{itemize}
\item \texttt{SN} contains the detection significance of a particular
  emission line.
\item \texttt{RA\_SN}, \texttt{DEC\_SN}, and \texttt{LAMBDA\_SN}
  contain the 3D S/N weighted right ascension, declination, and
  wavelength coordinate.
\item \texttt{RA\_PEAK\_SN}, \texttt{DEC\_PEAK\_SN}, and
  \texttt{LAMBDA\_PEAK\_SN} contain the position of the emission line
  peak in the S/N cubes.
\item \texttt{RA\_1MOM}, and \texttt{DEC\_1MOM} contain the first
  central image moment coordinate determined in a synthesised narrow
  band image of the emission lines (cf.
  Sect.~\ref{sec:line-source-param}).
\item \texttt{R\_KRON} contains the Kron-radius of a line detection
  determined in the same synthesised narrow-band (cf.
  Sect.~\ref{sec:line-source-param}).
\item \texttt{LAMBDA\_NB\_MIN} and \texttt{LAMBDA\_NB\_MAX} contain the
  maximum and minimum wavelength of the synthesised narrow band.
\item \texttt{F\_KRON}, \texttt{F\_2KRON}, \texttt{F\_3KRON}, and
  \texttt{F\_3KRON} contain the integrated flux over the emission
  line in the synthesised narrow-band within a circular aperture of
  radius $R_\mathrm{Kron}$, $2 \times R_\mathrm{Kron}$,
  $3 \times R_\mathrm{Kron}$, and $4 \times R_\mathrm{Kron}$
  respectively (see also Sect.~\ref{sec:line-source-param}).
\item \texttt{F\_KRON\_ERR}, \texttt{F\_2KRON\_ERR},
  \texttt{F\_3KRON\_ERR}, and \texttt{F\_3KRON\_ERR} contain the
  propagated errors on the flux measurements.
\end{itemize}
For algorithmic details on the above listed measurements we refer the
reader to the \texttt{LSDCat} paper (HW17).

\subsection{1D Spectra}
\label{sec:1d-spectra}

We provide a 1D spectrum for each of our objects.  These are created
by utilising a 2D weighted extraction in each spectral layer.  As
weights we adopt a normalised 2D Gaussian PSF, where we use the same
parameterisation that we used for the matched-filtering explained in
Sect.~\ref{sec:cross-corr-with}.  For compact, unresolved sources, the
spectral pixels of the so extracted 1D spectra give a nearly unbiased
and S/N-optimised estimate of the total flux within each spectral
layer.  However, for extended sources (or for extended emission line
profiles in compact sources), these spectra underestimate the total
flux.  These spectra should therefore only be used to get a first
impression of the spectral characteristics of a source in our
catalogue, while measurements should be performed on the source
datacubes described in Sect.~\ref{sec:source-datacubes} below.

We distribute the 1D spectra as FITS binary tables with file names
\texttt{spectrum\_ABBCCCDDD.fits}, where \texttt{ABBCCCDDD} is the
unique ID introduced in Sect.~\ref{sec:object-table}.  The tables
contain four columns: \texttt{AIR\_WAVE} and 
\texttt{AIR\_VAC} contain the air- and vacuum wavelength,
respectively, in \AA{}ngstrom (see note below Table~\ref{tab:emlines}
for details on the applied conversion).  \texttt{FLUX} and
\texttt{FLUX\_ERROR} contain the flux and the propagated error
in the extraction in units of
$10^{-20}$\,erg\,s$^{-1}$cm$^{-2}$\AA{}$^{-1}$.

\subsection{Source datacubes}
\label{sec:source-datacubes}

For each emission line galaxy we also provide three-dimensional source
datacubes.  These source datacubes are created from the
per-pointing datacubes whose reduction was described in
Sect.~\ref{sec:obs-reduc-musewide}.  Each source datacube is centred
on the coordinate given in column \texttt{RA} and \texttt{DEC} of the
object table (Sect.~\ref{sec:object-table}).  The spatial dimensions
of the extracted cubes are set to encompass 4.5 times the maximum
Kron-radius from the set of an object's detected emission lines.  We
do not remove or mask neighbouring objects that fall into the
extracted field of view of the source datacubes.  Moreover, if objects are
close to the border of the FoV, the  source datacubes contain voxels set
to \texttt{nan} (not a number) in regions where no data were
available.

We distribute these source datacubes as FITS files named
\texttt{objcube\_ABBCCCDDD.fits}, where \texttt{ABBCCCDDD} is the
unique ID introduced in Sect.~\ref{sec:object-table}.  These FITS
files contain three HDUs.  The first HDU stores the actual minicube in
a 3D flux array in units of
$10^{-20}$\,erg\,s$^{-1}$cm$^{-2}$\AA{}$^{-1}$, while the second HDU
stores the empirical noise estimate $\sigma_\mathrm{emp.}$
(Sect.~\ref{sec:effective-noise}) as an 1D array, in the same units,
for the pointing where the mini-cube was extracted.  

The third HDU of the mini-cube FITS file contains a 2D exposure map of
the same spatial dimensions as the mini-cube itself.  Each pixel of
this map shows the average number of exposures that went into the
corresponding spaxel.  Since each MUSE-Wide cube consists of four
exposures, the exposure map values range from zero to four.  Values
less than four are encountered either when single exposures were
affected by cosmic rays on the detector and thus was masked out in the
\texttt{muse\_scipost} resampling process
(Sect.~\ref{sec:obs-reduc-musewide}), or near the FoV borders where
several spaxels contain less than four exposures because of the
drizzling pattern.  Outside the FoV the values of the exposure map are
zero.

\section{Characteristics of the sample}
\label{sec:char-catal}

While an extensive scientific exploration of the presented sample is
beyond the scope of this paper, we provide in this section some basic
characteristics of our emission line selected galaxy sample from the
first 24 MUSE-Wide pointings.

\subsection{Redshift distribution}
\label{sec:redsh-distr}

\begin{figure*}
  \centering
  \includegraphics[width=0.45\textwidth]{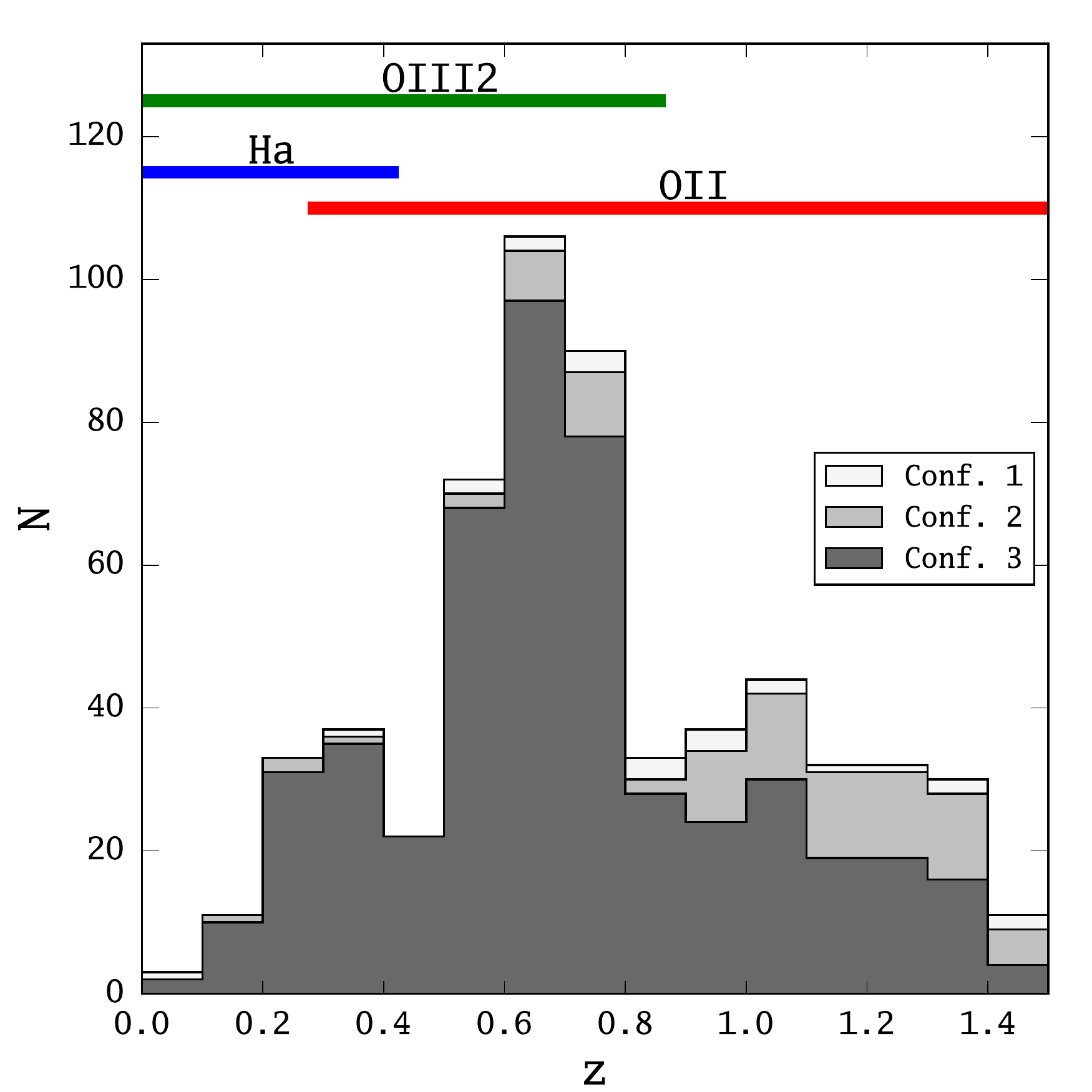}
  \includegraphics[width=0.45\textwidth]{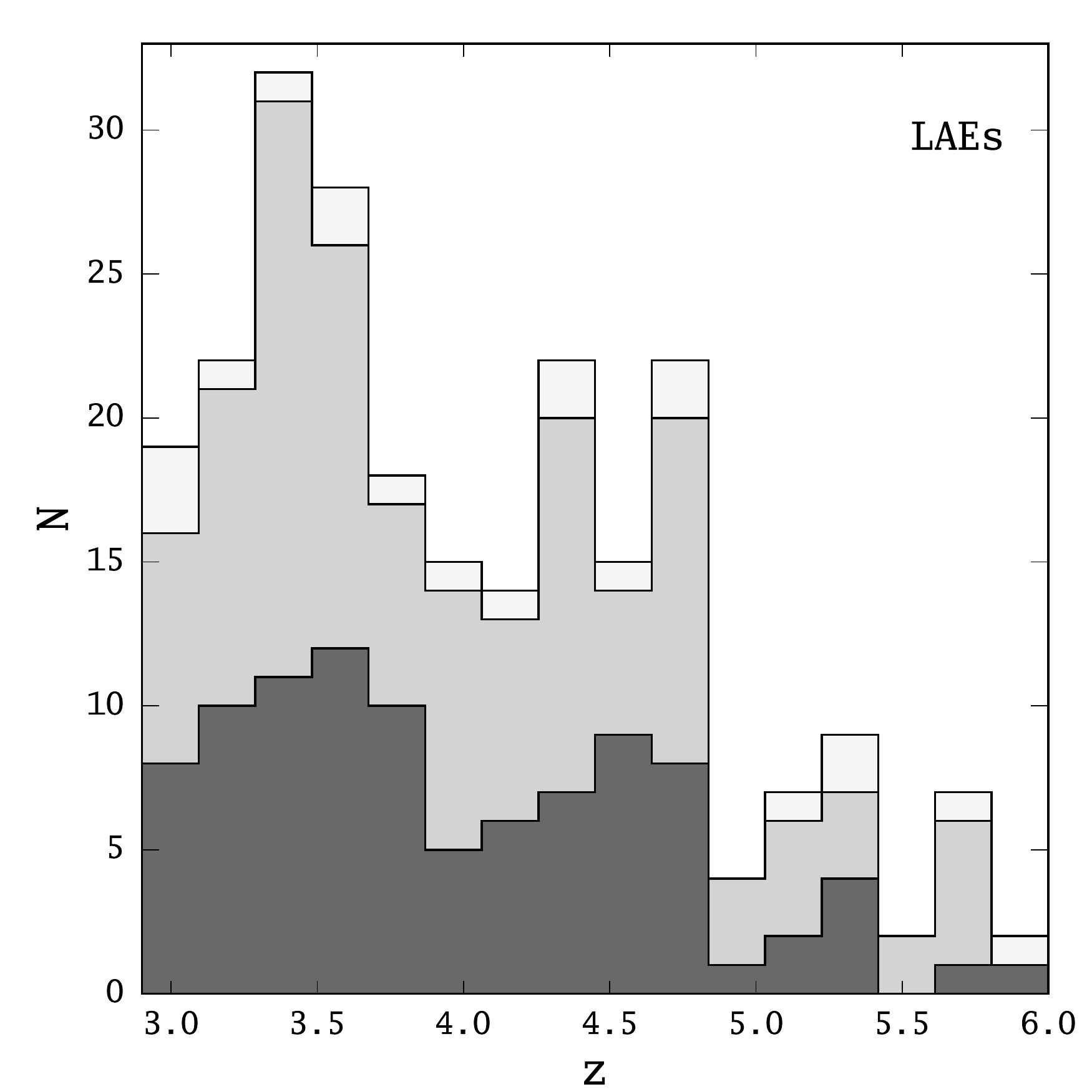}
  \caption{Histograms showing the redshift distribution of the 831
    emission line selected galaxies from the first year of MUSE-Wide
    observations.  Redshifts with confidence values 3, 2, and 1 are
    shown in dark grey, grey, and light grey, respectively.
    \emph{Left panel}: Redshift distribution for the rest-frame
    optical emission line selected galaxies. Horizontal bars indicate
    the redshift range of the MUSE wavelength coverage for the
    strongest emission lines of star-forming galaxies.  The bin size
    is $\Delta z = 0.1$. \emph{Right panel}: Redshift distribution for
    the 238 high-z galaxies. The bin size is $\Delta z = 0.2$.}
  \label{fig:redshift_dist}
\end{figure*}

In Fig.~\ref{fig:redshift_dist} we show a histogram of the
redshift-distribution obtained from the emission line fits described
in Sect.~\ref{sec:redshift}.  We have 595 galaxies at $z<2$ detected
by their rest-frame optical emission lines and 238 $z>2.95$ galaxies,
of which 237 where detected by strong Ly$\alpha$ emission and a single
object where the \ion{C}{IV} had a higher S/N than Ly$\alpha$.  At the
depth of MUSE-Wide the absence of strong nebular lines between
[\ion{O}{ii}] and Ly$\alpha$ results in a ``redshift desert''
in the interval $1.5\lesssim z \lesssim 3$.

\subsection{Redshift comparison with existing photometric and spectroscopic catalogues}
\label{sec:comparision-with-3d}

\begin{figure*}
  \centering%\vspace{-2em}
  \includegraphics[width=0.75\textwidth,trim=0 2em 0 6em,clip]{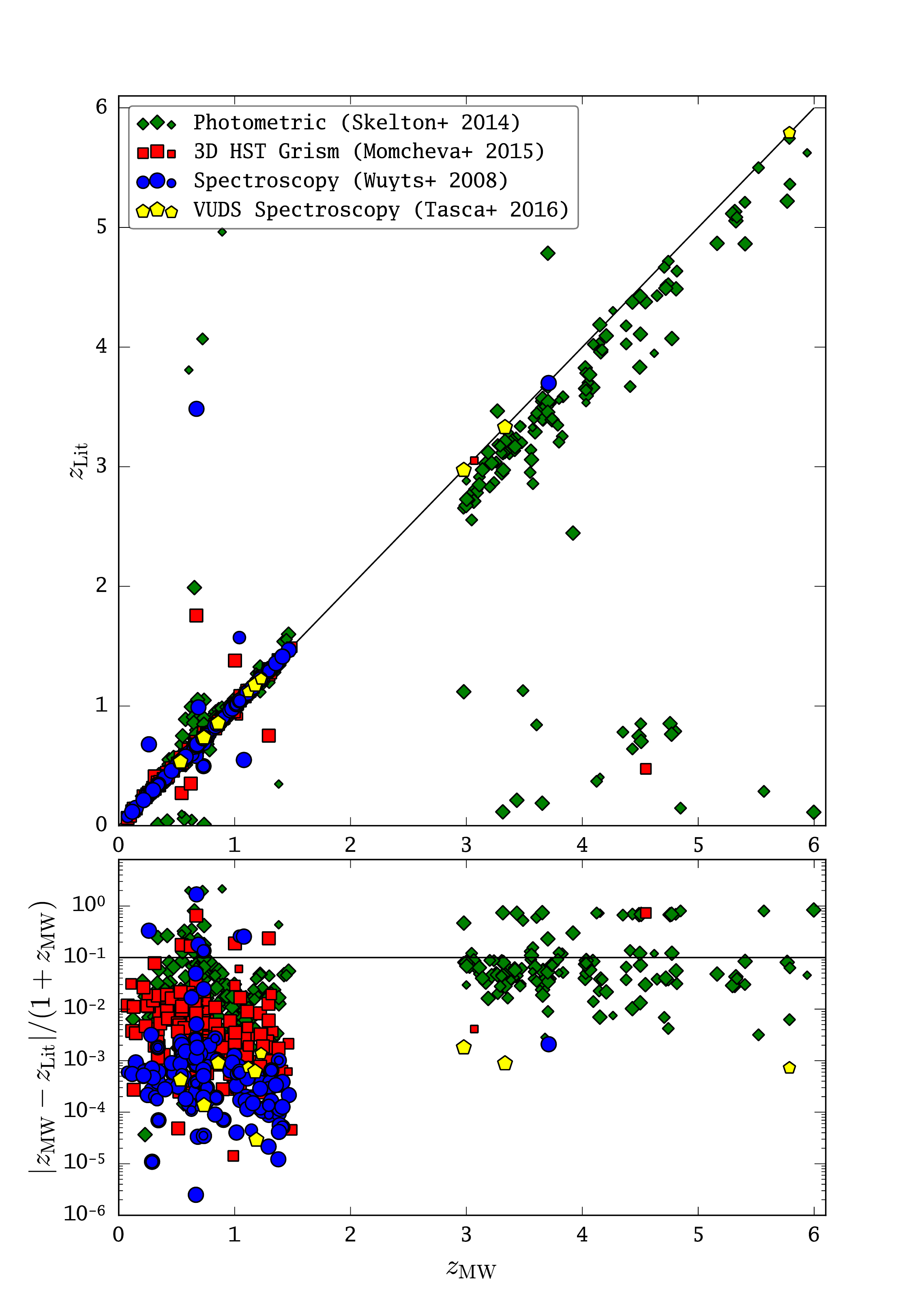}%\vspace{-2em}
  \caption{Redshift comparison between our MUSE-Wide emission line
    selected galaxy sample in the CDFS and literature redshifts (see
    text for cross-matching details). Literature redshifts are from
    the catalogues from \cite{Momcheva2016} and \cite{Tasca2016},
    where the former includes a compilation of ground based
    spectroscopic redshifts from \cite{Wuyts2008} and
    photometric-redshifts from \cite{Skelton2014}. See text for
    further details on the cross-matching procedure.  \emph{Top
      panel}: Literature redshifts ($z_\mathrm{Lit}$) versus MUSE-Wide
    redshifts ($z_\mathrm{MW}$). Symbol size encodes the confidence on
    the MUSE-Wide emission line classification
    (Sect.~\ref{sec:classification}), with the largest symbols
    representing ``Confidence 3'' sources, the medium sized symbols
    representing ``Confidence 2'' sources, and the smallest symbols
    representing ``Confidence 1''.  The different symbols encode the
    source of the catalogue redshifts according to the legend.
    \emph{Bottom panel:} Absolute relative difference between
    literature and MUSE-Wide redshifts.  The horizontal line denotes
    the boundary
    $|z_\mathrm{MW} - z_\mathrm{Lit}|/(1 + z_\mathrm{MW}) = 0.1$ where
    we distinguish between redshift equality and redshift mismatches.}
  \label{fig:zcomp}
\end{figure*}

\begin{table}
  \centering
  \caption{Catastrophic redshift mismatches (defined as
    $|z_\mathrm{MW} - z_\mathrm{Lit}|/(1 + z_\mathrm{MW}) > 0.1$)
    between MUSE-Wide and literature redshifts for objects within a
    0.5\arcsec{} search radius around a MUSE-Wide position.}
    \label{tab:redshiftmiss}
    \begin{tabular}{l@{}cc} \hline \hline Redshift source & $z\lesssim 2$ &
      $z \gtrsim 3$ \\ \hline
      Spectroscopy \citep{Wuyts2008}      & 7/139 (5\%) & 0/2  (0\%)  \\
      Spectroscopy \citep{Tasca2016} & 0/7 (0\%) & 0/3  (0\%)  \\
      3D-HST grism \citep{Momcheva2016}  \;    & 6/182  (3\%) & 1/2  (50\%) \\
      Photometric \citep{Skelton2014} & 19/241 (8\%) & 31/103 (30\%) \\ \hline total & 32/569 (5\%) & 32/110 (29\%)
      \\ \hline
    \end{tabular}
    \tablefoot{In the form $x/y$, where $x$ is the number of redshift
      mismatches and $y$ the total number of secure counterparts with literature redshifts.
      In brackets this fraction is expressed as percent.
    }
\end{table}

\begin{table*} 
\centering
\caption{Crossmatch between MUSE emission line galaxies and X-ray
  sources flagged as AGN from the {\it Chandra} 7Ms source catalogue
  \citep{Luo2017}}
\begin{tabular}{c c c c c c}
  \hline \hline
ID & $z$ & ID  & Separation & X-ray Flux  & $z$ \\
MUSE Wide & MUSE & Chandra 7Ms & (``) & (erg/s/cm$^2$) & Chandra \\
\hline
102007068  & 0.338 & 304 & 0.20 & 2.982e-16 & 0.340 \\       
102008071  & 0.338 & 312 & 0.86 & 6.491e-17 & 0.336 \\   
102031144  & 0.665 & 290 & 0.20 & 3.498e-16 & 0.664 \\       
102037154  & 1.412 & 287 & 0.51 & 5.925e-17 & 1.413 \\       
103022086  & 0.670 & 322 & 0.24 & 1.510e-16 & 0.671 \\      
104014050  & 3.662 & 337 & 0.17 & 1.269e-15 & 3.660 \\      
106036089  & 0.904 & 344 & 0.82 & 6.638e-17 & 0.956$^a$ \\     
106048103  & 0.665 & 340 & 0.15 & 2.406e-15 & 0.666 \\      
108025145  & 0.736 & 407 & 0.49 & 7.387e-17 & 0.736 \\    
109030090  & 1.044 & 447 & 0.17 & 5.737e-16 & 1.043 \\     
111004005  & 0.604 & 367 & 0.28 & 4.594e-15 & 0.604 \\    
113001007  & 0.232 & 508 & 0.94 & 2.411e-17 & 0.220$^a$ \\      
113010038  & 0.577 & 460 & 0.35 & 2.778e-16 & 0.577 \\    
114024110  & 1.035 & 443 & 0.26 & 8.709e-16 & 1.036 \\       
114028115  & 1.098 & 509 & 0.30 & 1.525e-16 & 1.097 \\       
115003085  & 3.710 & 551 & 0.11 & 2.158e-15 & 3.700 \\      
116003060  & 1.364 & 634 & 0.14 & 3.899e-17 & 1.363$^b$ \\     
117034085  & 0.228 & 693 & 1.13 & 6.435e-17 & 2.302 \\     
119034073  & 1.015 & 814 & 0.15 & 4.348e-15 & 1.016 \\    
120023032  & 1.118 & 861 & 0.25 & 1.744e-16 & 1.120 \\    
123005089  & 0.544 & 640 & 0.51 & 4.219e-17 & 0.552$^b$ \\   
123051191  & 4.510 & 625 & 1.33 & 2.102e-17 & 2.616$^b$ \\    
  \hline
\end{tabular}
\tablefoot{All reshifts quoted are spectroscopic from extensive
  identification campaigns except for objects marked with $^a$ and
  $^b$ which are photometric redshifts from \cite{Skelton2014} or
  \cite{Hsu2014}, respectively.}
\label{tab:AGNtable}
\end{table*}

We now compare our 831 emission line selected galaxies to existing
photometric and spectroscopic redshift measurements from  the
literature within our survey area.

\cite{Momcheva2016} presented a redshift catalogue of $\sim10^5$
sources in all CANDELS fields containing grism redshifts from the
3D-HST survey \citep{Brammer2012,Momcheva2016}, photometric redshifts
from 3D-HST/CANDELS photometry \citep{Skelton2014}, as well as a
compilation of ground-based spectroscopic redshifts also contained in
\cite{Skelton2014}.  For the CDF-S region under scrutiny here the
ground-based redshifts in this compilation are taken from the
compilation by \citeauthor{Wuyts2008} (\citeyear{Wuyts2008}; see their
Table 3 for individual spectroscopic campaigns).
To be conservative, we here include only objects that have a
photometric counterpart within a 0.5\arcsec{} radius of the MUSE-Wide
position, i.e. we discard all manually assigned cross-matches
from Sect.~\ref{sec:object-table}.  141 of the remaining 669 secure
associations have a ground based spectroscopic redshift (139 at
$z\lesssim 2$ and 2 at $z\gtrsim3$), 184 have a grism redshift (182 at
$z\lesssim 2$ and 2 at $z\gtrsim 3$), and 344 have a photometric
redshift (241 at $z\lesssim 2$ and 103 at $z\gtrsim 3$).

Using the search radius of 0.5\arcsec{} we also cross-matched our
objects with the first data release of the VIMOS Ultra Deep Survey
\citep{LeFevre2015,Tasca2016}. We find matches to 10 of our objects in their
spectroscopic redshift catalogue: seven at $z<1.5$, and three at $z\gtrsim 3$
galaxies.

In Fig.~\ref{fig:zcomp} we compare the literature redshifts for the
679 objects that could be cross-matched with the 3D-HST or VUDS
catalogues.  Following the literature \citep[e.g.][]{Skelton2014}, we
define a catastrophic redshift mismatch between two catalogues as one
with $|z_\mathrm{MW} - z_\mathrm{Lit}|/(1+z_\mathrm{MW}) > 10$ \%{},
where $z_\mathrm{MW}$ is the MUSE-Wide redshift and $z_\mathrm{Lit}$
is the corresponding literature redshift.  Based on this definition 64
of the 3D-HST redshifts (i.e., 9.6\% of secure cross-matches within a
search radius 0.5\arcsec{}) are in disagreement.  On the other hand
the 10 VUDS redshifts are in excellent agreement with the MUSE-Wide
redshift, all characterised by
$|z_\mathrm{MW} - z_\mathrm{VUDS}|/(1+z_\mathrm{MW}) < 10^{-3}$.

In Table~\ref{tab:redshiftmiss} we summarise the statistics on the
redshift mismatches and indicate whether the mismatch is a low- or
high-$z$ source.  We find the highest rate of catastrophic mismatches
(30\%{}) between MUSE-Wide and literature redshifts amongst
photometrically determined redshifts at $z\gtrsim 3$.  For the 3D-HST
grim redshifts the agreement is mostly excellent, except for a few
cases at low-redshift and one of the only two grism redshift at
high-$z$.  Notably, all low-$z$ mismatches have the highest confidence
value in our catalogue, meaning that the MUSE-Wide data leaves no
doubt on the obtained redshift. The high-$z$ grism redshift mismatch
could be reconciled if we would classify this single emission line
detection as [\ion{O}{ii}] instead of Ly$\alpha$.  Based on our visual
inspection of the line profile, however, we flagged this source with
confidence\,$=2$, meaning that we have at most minor doubts on our line
classification.  Finally, only 5 from 140 spectroscopic redshifts from
\cite{Wuyts2008} are in disagreement with our catalogue.  Four of
those were marked with the highest confidence flag in our
classification and only in one case we did have minor doubts.  Notably, one
single line detection classified as Ly$\alpha$ in the \cite{Wuyts2008}
compilation clearly shows a characteristic \ion{O}{ii} profile that
likely was not resolved in the previous spectrum.

We notice that for high-$z$ LAEs in our sample a systematic offset
between photometric and spectroscopic redshifts with
$|z_\mathrm{MW} - z_\mathrm{3DHST}|/(1+z_\mathrm{MW}) \sim 10^{-1.5}$
exists. 23\% of the LAEs have
$\Delta z = z_\mathrm{MW} - z_\mathrm{Lit} > +0.1$ and the median
$\Delta z$ is +0.26, with the upper and lower quartiles being +0.17
and +0.44, respectively.  This systematic offset is much larger then
the known systematic offset of the Ly$\alpha$ line to the systemic
redshift discussed in Sect.~\ref{sec:high-redshift-lyman}.  Recently,
\cite{Oyarzun2016} reported very similar offsets between photometric
and spectroscopic redshifts of LAEs.  They found that the magnitude of
the offset correlates with the Ly$\alpha$ equivalent width.  Hence, a
possible source of the discrepancy is that Ly$\alpha$ line flux is
missing from the spectral energy distribution templates which are used
in the photometric redshift fitting.

\subsection{Emission line fluxes and continuum magnitudes}
\label{sec:emission-line-fluxes}

\begin{figure*}
  \centering
  \includegraphics[width=\textwidth,trim=0 0 5 0,clip=true]{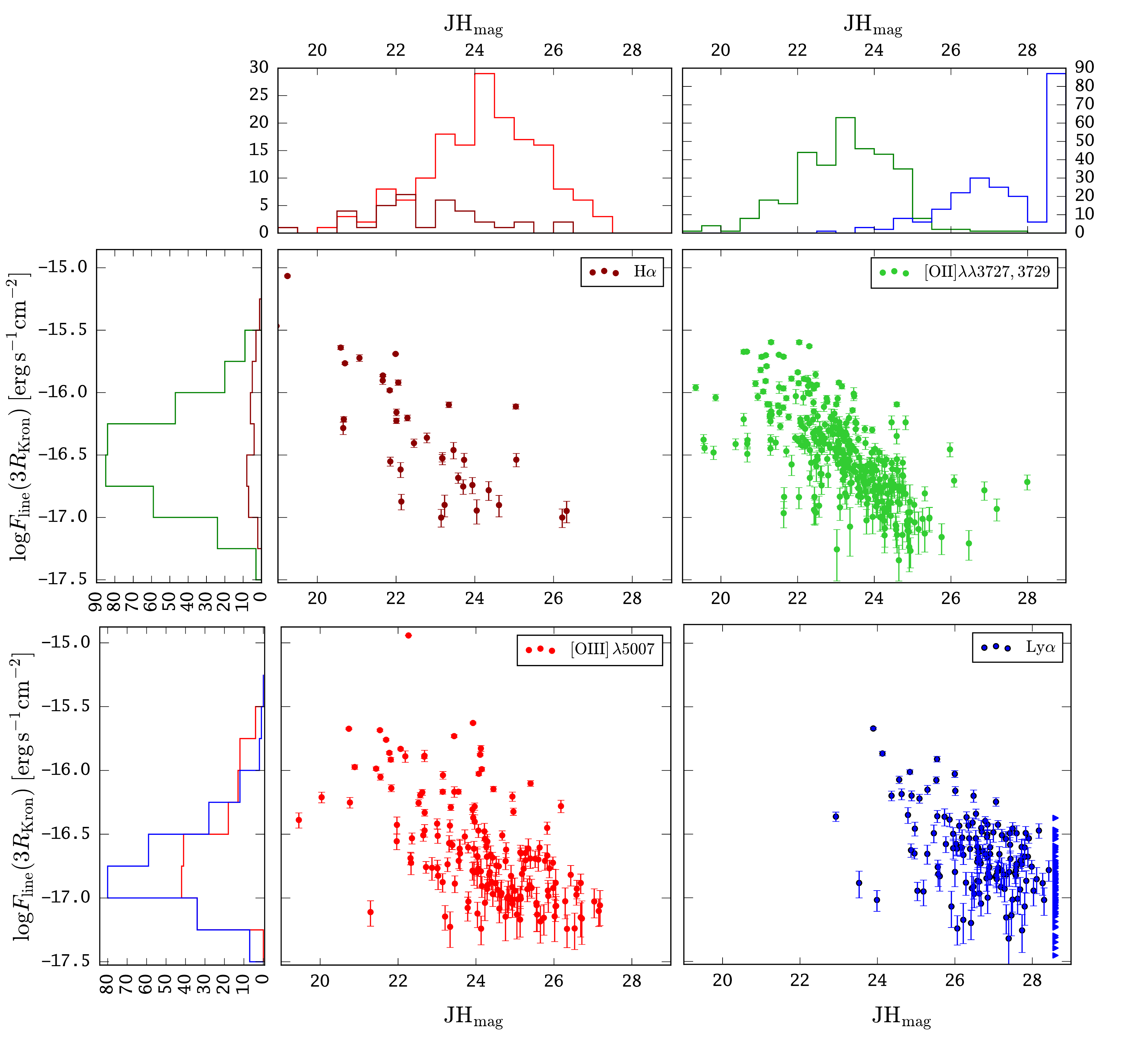}\vspace{-1em}
  \caption{Line fluxes of the strongest emission lines vs. continuum
    $JH_\mathrm{IR}$ magnitudes from 3D-HST \citep{Momcheva2016} for
    669 of 831 MUSE-Wide galaxies that have a 3D-HST catalogue
    counterpart within 0.5\arcsec{}.  H$\alpha$, [\ion{O}{ii}],
    [\ion{O}{iii}], and Ly$\alpha$ lines are shown in brown, green,
    red, and blue, respectively. With the same colours we also show
    histograms of the scatter-clouds.  Ly$\alpha$ objects not having a
    counterpart within $r=0.5$\arcsec{} in the 3D-HST catalogue are
    shown as blue triangles and have been put into the
    $JH_\mathrm{IR}=29$\,mag bin.  We point out, however, that most of
    the objects show photometric counterparts that are not present in
    the catalogue. }
  \label{fig:swf}
\end{figure*}

To characterise our sample we compare in Fig.~\ref{fig:swf} the
measured line fluxes in 3$\times R_\mathrm{Kron}$ apertures with the
continuum magnitudes in the $JH_\mathrm{IR}$ magnitude from the 3D-HST
catalogues.  As described in Sect.~3.6 of \cite{Momcheva2016} (see
also their Table~5) the $JH_\mathrm{IR}$ magnitudes have been measured
with SExtractor's \texttt{MAG\_AUTO} in a coadded image created from
the HST WFC3 $J_{125}$, $JH_{140}$, and $H_{160}$ images.

Redshifts from the 3D-HST grism survey are provided for objects with
$JH\leq24$.  Similarly, previous ground based spectroscopic deep-field
follow-up campaigns target objects to a limiting photometric depth.
For example, the VIMOS VLT Deep sample contains objects with
$i_\mathrm{AB}$-magnitudes brighter than 24.75 \citep{LeFevre2013}.
However, as can be seen from Fig.~\ref{fig:swf}, a significant
fraction of galaxies in MUSE-Wide is characterised by very faint
continuum emission.  This is especially so for the Ly$\alpha$ emitting
high-$z$ galaxy population in our catalogue and thus explains the
previous lack of spectroscopic redshifts for those galaxies.

The $JH_\mathrm{IR}$ distribution of tha LAE population peaks at
$JH_\mathrm{IR}\sim27$ for objects that have a photometric counterpart
in the CANDELS 3D-HST catalogue.  However, 87 MUSE-Wide LAEs have no
cross-matches to the deep 3D-HST catalogue within 0.5\arcsec{}. We
place those objects arbitrarily at $JH_\mathrm{IR}=29$ in
Fig.~\ref{fig:swf}.  We caution that most of these objects in fact
show prominent photometric counterparts but were missed in the 3D-HST
NIR-based detection.  Therefore we expect most of the upper limits in
the bottom right panel of Fig.~\ref{fig:swf} moving towards brighter
magnitudes.  A detailed evaluation of the continuum-faint population
of MUSE-Wide LAEs is beyond the scope of this catalogue release and
will be subject of a future study.

\subsection{Active Galactic Nuclei}
\label{sec:active-galaxies}

Traditionally blind spectroscopic surveys hunting for emission lines
galaxies  unveiled numerous active galaxies \citep[e.g.][]{Zamorano1994},
so naturally we want to check that in our sample as well. At low
redshift, where H$\alpha$ and nearby lines fall within the wavelength
range covered by {\it MUSE}, we can employ the BPT diagram \citep{Baldwin1981} 
to distinguish AGN from star forming galaxies. However, for larger
redshifts, we need to rely on ancillary spectroscopic data from the
near-IR to use the BPT diagram. This is beyond the scope of this
paper. 

We therefore opt to crossmatch our emission line sources with the main
catalog of the 7Ms Chandra Deep Field South \citep{Luo2017}. A match
was deemed succesful if our emission line position was within 3 times
the X-ray positional accuracy (SIGMAX or column 6 in Table 4 of the
\citep{Luo2017} catalogue). In addition to an X-ray match, we required
that the object type column (column 70) of the X-ray catalog be
``AGN'' as to not include any X-rays from star formation regions. We
find 22 candidate AGNs in our emission line catalog which we present
in Table \ref{tab:AGNtable}.

Most of the matched sources already have been spectroscopically
identified in previous campaigns to identify optical counterparts to
X-ray sources (e.g. \cite{Szokoly2004}, see \cite{Luo2017} for an
extensive list of 26 identification references). Nevertheless, we can
assign spectroscopic redshifts to 5 objects which previously only had
photometric redshifts determined. All, but two of the redshifts are in
excellent with the MUSE determined redshift. These are the ones with
the largest distance between the X-ray and MUSE position, so most
likely a different X-ray counterpart was assigned. Among the 17 AGN
are two well known type 2 high redshift quasars (MUSE-WIDE IDs:
104014050, 115003085), both of which have the highest Ly$\alpha$ flux
in our survey \citep{Norman2002,Mainieri2005}.

\section{Conclusions and outlook}
\label{sec:musewide-samp-conc}

We present the first results from the ongoing MUSE-Wide survey. Using
LSDCat, a novel 3D source detection algorithm based on matched
filtering, we constructed a catalogue of 831 galaxies with altogether
1656 detected emission lines, all located in a footprint of 22.2
arcmin$^2$ within the CANDELS-Deep/CDFS region. More than half of
these galaxies did not have spectroscopic or HST grism redshifts until
now.

Because of the emission line selection, the properties of these
galaxies are quite different from those of photometrically preselected
samples: The redshift distribution features two disjoint domains, a
low to intermediate redshift category with $z<1.5$ where rest-frame
optical nebular lines are detected, and a high-redshift ($z>2.9$)
category of strong Ly$\alpha$ emitters; in between these domains,
MUSE-Wide suffers from the well-known ``redshift desert'' effect
caused by the dearth of strong emission lines in the UV.
Nevertheless, compared to the already extensive spectroscopic coverage
in these fields, MUSE-Wide adds significantly to both the
low/intermediate- and the high-redshift domain by achieving a spatial
target sampling rate of essentially 100\%, irrespective of the
distribution and shapes of sources in the sky.

While our adopted selection criteria reveal objects that are similar
to those found in narrowband imaging, there are also some notable
differences. Rather than by covering a large solid angle in the sky,
at predefined small redshift windows, such as obtained by instruments
like Hyper Suprime-Cam \citep{Miyazaki2012}, MUSE-Wide gains its survey
volume through its high redshift path length. Since all our survey
fields are fully included within the footprints of very deep
multi-band HST imaging and other multi-wavelength deep field efforts,
the amount of available information per object is maximised. An
obvious further advantage over narrowband-selected samples is the fact
that all MUSE-Wide sources are already spectroscopically confirmed.

In order to promote the legacy aspect of our survey we provide not
only the spectra, catalogue data, and flux measurements, but also
datacube cutouts centred on each of the 831 objects. Several of the
low and intermediate-redshift objects in our sample are clearly
spatially extended. With HST prior morphological information and using
our datacube cutouts it will be possible, e.g., to perform spatially
resolved kinematic analyses \citep[such as in][]{Contini2016} or
investigate gas-phase metallicity gradients, thus conduct studies that
normally come at considerable extra observational costs.

Of particular scientific interest is the high-redshift part of this
sample, because of the unique combination of MUSE IFU with deep HST
data. Remarkably, only 3 out of the 238 high-$z$ emission line objects
in the surveyed 22.2 arcmin$^2$ were previously known as
spectroscopically confirmed Ly$\alpha$ emitters. It is now established
that Ly$\alpha$ in high-$z$ galaxies essentially always has a
significant circumgalactic component that extends much beyond the
stellar continuum emission region \citep{Wisotzki2015}.  MUSE now
facilitates the inclusion of this extended nature of the Ly$\alpha$
line into demographic studies, especially important for quantifying
the completeness of LAE selection and the impact on the Ly$\alpha$
luminosity function (Herenz et al., in prep.). We are also
investigating several other aspects of LAE demographics which will be
presented in a suite of forthcoming papers.

The present sample is by construction restricted to galaxies with
significantly detected emission lines. As a next step we are working
on an extended redshift survey that includes also continuum-selected
objects with only weak or without emission lines. This extension will
further enhance the legacy value of MUSE-Wide, especially for
relatively low redshifts. Furthermore, results from the deep and
ultra-deep MUSE surveys (\citealt{Bacon2015} and in prep.) will provide
constraints on the faintest parts of the galaxy population at all
redshifts accessible to MUSE. Combining such multi-tier approaches has
always been a winning strategy for astronomical surveys, and MUSE
makes no exception here.

At the time of writing, observations for MUSE-Wide are proceeding well
and the goal of observing 100 MUSE pointings was reached February
2017. Besides serving our own scientific interests, this survey will
remain a very valuable community resource, and we have committed
ourselves to release the data to the public as soon as possible, after
careful quality control and postprocessing where needed. The first
comprehensive data release (DR1, Urrutia et al. in prep) is currently
anticipated for 2017, covering 44 MUSE fields in the CDFS
region. Further data releases will follow.

\begin{acknowledgements}
  We thank the support staff at ESOs VLT for help with the visitor
  mode observations.  This research made extensive use of the
  \texttt{astropy} pacakge \citep{AstropyCollaboration2013}. All plots
  in this paper (except Fig.~\ref{fig:qtclassify}) were created using
  \texttt{matplotlib} \citep{Hunter2007}. E.C.H, J.C., J.K., R.S.,
  T.U., and L.W. acknowledge funding by the Competitive Fund of the
  Leibniz Association through grants SAW-2013-AIP-4 and
  SAW-2015-AIP-2.  R.B. acknowledges support from the ERC advanced
  grant 339659-MUSICOS.
\end{acknowledgements}

\appendix

% References
\bibliographystyle{aa}
\bibliography{mw_sample_paper.bib}

\newpage

\section{QtClassify}
\label{sec:qtclassify}

\begin{figure*}
  \centering
  \includegraphics[width=\textwidth]{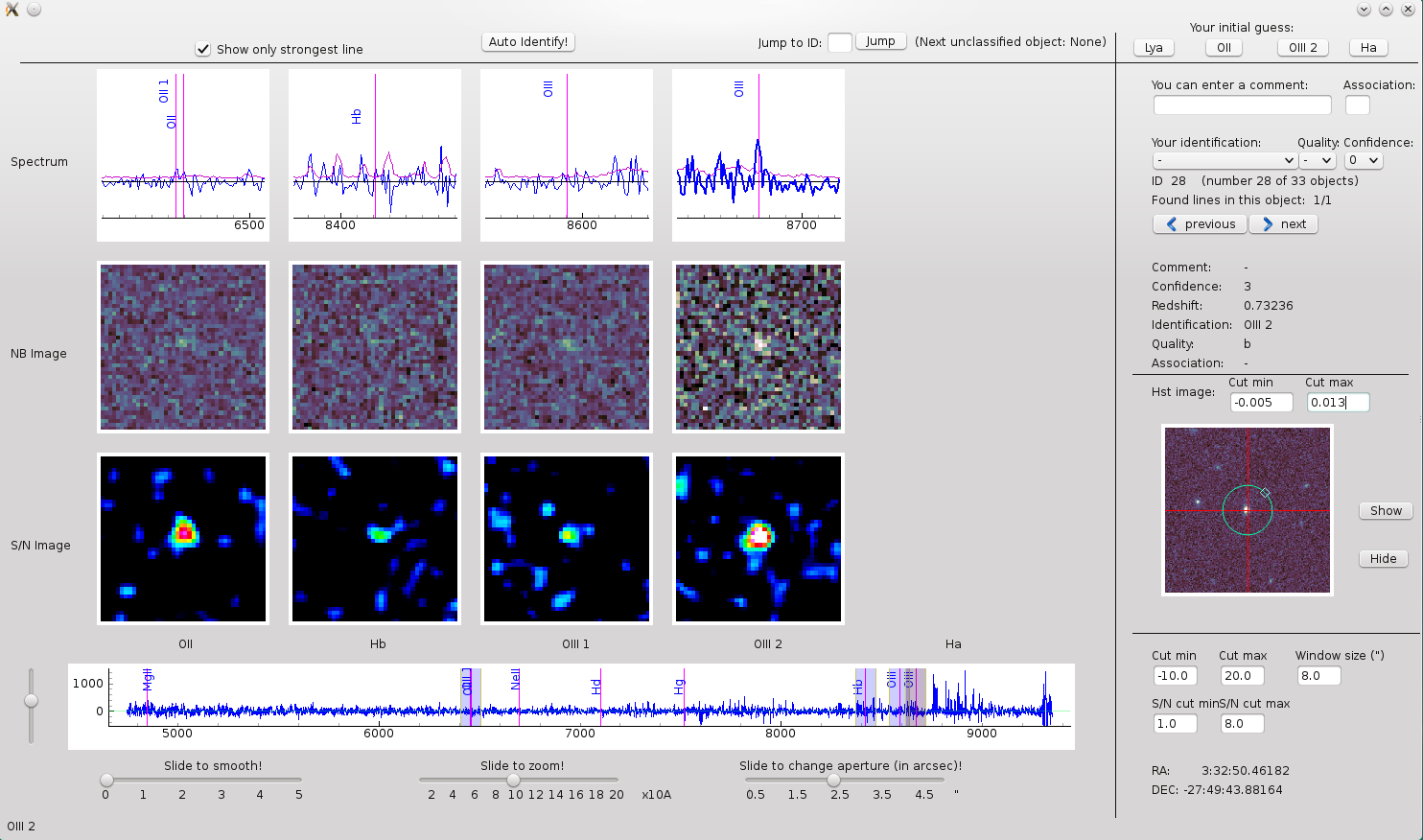}
  \caption{Screenshot of QtClassify, the software software for
    classification of emission line sources detected in wide-field IFS
    datacubes \citep{Kerutt2017}.  All graphical panels are
    interactive.  The top row panels display sections of a spectrum
    extracted from the datacube at the position of the detected
    emission line under scrutiny.  The aperture radius for extraction,
    the limits of the y-axis, as well as the degree of smoothing of
    the spectrum can be changed interactively using the sliders at the
    bottom.  The positions of displayed sections are adjusted to show
    regions around possible veto lines under the users guess for the
    nature of the detected emission line.  Below the spectra
    pseudo-narrow band images at the veto line positions are
    displayed.  Here the cut-levels as well as the width of the
    narrow-band window can be altered interactively at the bottom
    right section of the GUI.  The bottom row of panels displays the
    corresponding layers from \texttt{LSDCat}'s S/N-cube.  Both the
    spectral position of the center-row panels and the bottom-row
    panels can be changed simultaneously with a slider or by moving
    the pointer in the top-row panels. On the bottom of the window an
    overview over the whole extracted spectrum is displayed, and at
    the right side of the window a cut-out from 2D imaging data can be
    shown.  The example shown here demonstrates the classification of
    a rather weak [\ion{O}{iii}] emission line. For this object other
    lines (\ion{O}{ii}, H$\beta$, and [\ion{O}{iii}] $\lambda$4959)
    are below the detection threshold, but can still be visually
    identified in the different panels of the window.}
  \label{fig:qtclassify}
\end{figure*}

QtClassify is a graphical user interface (GUI) that helps to classify
emission line objects detected in wide-field IFS data. Its development
was loosely inspired by the \texttt{SpecPRO} GUI used in classical
imaging and spectroscopic surveys \citep{Masters2011}.  QtClassify is
publicly available and can be downloaded via the Astrophysics Source
Code Library \citep{Kerutt2017}: \url{http://ascl.net/1703.011}.

In Fig.~\ref{fig:qtclassify} we display a screenshot of QtClassify.
As input it requires a -- preferentially continuum-subtracted --
datacube, a catalogue tabulating the positions of the emission lines,
and a S/N datacube.  Optionally, broadband imaging data can also be
loaded into the software.  For creating the catalogue and S/N cube
qtClassify is currently geared towards the outputs of \texttt{LSDCat}
\citep{Herenz2017}.  However, in principle it can be used with any IFS
datacube and any catalogue that contians one entry for each emission
line with all entries grouped into objects by position.

The main idea of QtClassify is to use the full 3D information in the
datacubes to classify emission lines by simultaneously displaying
regions in the cube where other lines would be expected given an
initial guess for the identification of the strongest detected line.
Strongest here means the line with the highest signal-to-noise ratio
(determined e.g. by the matched-filtering procedure in
\texttt{LSDCat}, but it is also possible to use line flux instead).
The default guesses are Ly$\alpha$ $\lambda$1216, [\ion{O}{ii}]
$\lambda\lambda$3726,3729, [\ion{O}{iii}] $\lambda$5007 and H$\alpha$
$\lambda$6563, but they be changed by customising the input list of
options.
 
To help classify the lines, there are several rows with information
(see Fig.~\ref{fig:qtclassify}). When clicking on one of the possible
guesses, the columns change to the positions where you would expect
other lines. The first row ('Spectrum') displays the part of the
automatically extracted spectrum with an indication where the line is
expected. The second row ('NB Image') shows a monochromatic layer of
the actual datacube and the third row ('S/N Image') shows a
monochromatic layer of the S/N cube.  When interacting with the
programme, the user can move the mouse through the spectral parts in
the first row, which will automatically also move through the
corresponding wavelength layers in the second and third rows
generating a ``movie-like'' sequence of monochromatic slices in the
datacube.  The bottom panel shows the full spectrum where the
zoomed-in parts of the first row are indicated by shaded areas.  On
the right of the window, a broadband image (in the example displayed
in Fig.~\ref{fig:qtclassify} from the HST CANDELS data) can be
displayed if needed.  The cuts for the broadband image as well as for
the monochromatic images of the datacube and the S/N cube can be
interactively adjusted.  Several sliders at the bottom of the window
allow interactive smoothing and zooming of the spectrum, as well as
changing the size of the extraction aperture for the displayed
spectrum.

% The latter is especially useful when trying to determine whether an
% emission line is Ly$\alpha$, since most Lyman alpha emitters (LAEs)
% have extended haloes, which increases the line flux when increasing
% the aperture.  This would not be the case for OII emitters for example,
% which can sometimes mimic the double peak of Lyman alpha lines. This
% makes it possible to distinguish between these two possibilities.
 
In order to make it less tedious to go through all the detected
emission lines, there is a button at the top that automatically
identifies all lines where a unique redshift solution can be
automatically anchored. For this, QtClassify looks at all objects with
multiple emission lines and internally tries to determine a redshift
using a list of typical emission lines \citep[similar to the peak
correlation algorithm by][]{Garilli2010}.  This list can be adjusted
and extended.  This automatic procedure fails, of course,
when only a single, maybe even spurious, emission line is detected, or
when a superposition of objects at different redshifts occurs at the
same position on the sky.  Those objects have to be manually
classified by the QtClassify user.
 
The interactive classification is done via a drop-down menu that
presents a list of emission lines to the user, including various
classes of spurious detections (e.g. cosmics, noise-peaks, telluric-,
or continuum-subtraction residuals).  In addition the classifier can
enter confidence value and quality flag
(cf. Sect.~\ref{sec:classification}).  Cases of overlaps, where one
entry in the catalogue has emission lines from multiple objects, are
not covered by QtClassify and have to be sorted out separately in a
later stage of the catalogue construction.  Here the comment function
enables one to take notes for this process.  Multiple entries at different
positions on the sky that belong to the same object can be marked as
associations.  It is possible to stop classifying at any point and
resume later with the output catalogue of QtClassify as input, since
it automatically saves the classifications in columns that are
appended to the input catalogue.

\end{document}